\documentclass[12pt,amsmath,amssymb,]{revtex4}
\usepackage{amsfonts}
\usepackage{amsmath}
\usepackage{amssymb}
\usepackage{array}
\usepackage{bm}
\usepackage{braket}
\usepackage{breqn}
\usepackage{color}
\usepackage{dcolumn}
\usepackage{epsfig}
\usepackage{epsf}
\usepackage{epstopdf}
\usepackage{float}
\usepackage{graphicx}
\usepackage{graphics}
\usepackage{harpoon}
\usepackage{indentfirst}
\usepackage{mathrsfs}
\usepackage{multirow}
\usepackage{ulem}

\newcommand {\sla}[1]{ #1 \!\!\!\!/}

\newcommand{\beq}{\begin{eqnarray}}
\newcommand{\eeq}{\end{eqnarray}}

\begin{document}

\title{$\gamma Z$-exchange contributions in  low-energy parity-violating $ep$ scattering}

\author{
Qian-Qian Guo and Hai-Qing Zhou\protect\footnotemark[1] \protect\footnotetext[1]{Email address: zhouhq@seu.edu.cn} \\
School of Physics,
Southeast University, NanJing 211189, China}

\date{\today}

\begin{abstract}
In this work, the $\gamma Z$-exchange contributions in the low-energy elastic parity-violating $ep$ scattering are discussed with the approximation $m_e=0$. By expanding the $\gamma pp$ and $Zpp$ interactions on the momentum of photon and considering both the leading-order and the next-to-leading order interactions, we calculate the amplitudes of the $\gamma Z$-exchange diagrams. After performing the loop integral, we expand the results in the low energy limit, and obtain the analytic expressions for the amplitudes. Numerical comparisons show that the analytic expressions are very close to the full results over a large region. We investigate the power behaviors of these contributions  and find that some are enhanced by a kinematic factor in the low energy limit. Additionally, in some cases, the imaginary parts of the contributions from the next-to-leading-order interactions are at the same order as those from the leading-order interactions. Furthermore, the corresponding contributions to the physical observable quantity $A_{\textrm{PV}}$ are also discussed. Combining all the properties together, we conclude that these analytic expressions describe the leading-order contributions of all $\gamma Z$-exchange helicity amplitudes in the region with $\alpha_e\ll Q/M_N\sim \delta/M_N^2\ll 1$, where $\alpha_{e}$ is the fine structure constant, $M_N$ is the mass of proton, $Q$ and $\delta$ are the small quantities related to the momentum transfer and the center-of-mass energy.
\end{abstract}

\maketitle


\section{Introduction}

The parity-violating effects in the elastic $ep$ scattering provide the way to extract the weak charge of proton $Q_{W}$ and the weak form factors of the proton $F_{1,2,3}^{(Zpp)}$. Typically, the measurement of the parity-violating asymmetry, defined as $A_{\textrm{PV}}\equiv (\sigma^+-\sigma^-)/(\sigma^++\sigma^-)$, is used to extract these quantities \cite{Kaplan1988,SAMPLE,HAPPEX,A4,G0,Qweak,P2}. To extract these physical quantities precisely, the virtual radiative and the real radiative corrections should be estimated carefully. Among the virtual radiative corrections, the contributions from the $\gamma Z$ exchange are particularly distinct, as their effects cannot be absorbed by certain constants even when the momentum transfer is fixed. In the literature, several methods have been employed to estimate the $\gamma Z$-exchange contributions to $A_{\textrm{PV}}$. These include traditional calculation with zero energy approximation \cite{ep-ep-gammaZ-zero-energy}, hadronic model \cite{ep-ep-gammaZ-hadronic-model-method}, general partonic distributions (GPDs) \cite{ep-ep-gammaZ-GPD-method}, and dispersion relations (DRs) method \cite{ep-ep-gammaZ-dispersion-relation-1, ep-ep-gammaZ-dispersion-relation-2}.

In this work, we discuss the $\gamma Z$-exchange contributions from a different perspective. We use the low-energy $\gamma pp$ and $Zpp$ interactions, which are expanded on the momentum to the leading-order (LO) and the next-to-leading-order (NLO), to calculate the $\gamma Z$-exchange amplitudes. A similar method has been used to discuss the two-photon-exchange (TPE) contributions in elastic $ep$ and $\mu p$ scattering \cite{lp-lp-TPE-chiral}. In this work, we provide the analytic expressions for the $\gamma Z$-exchange contributions at the amplitude level in the low energy limit. These expressions reveal several interesting properties, which are not readily apparent in the direct numerical results or the conventional estimation of the $\gamma Z$-exchange contributions to $A_{\textrm{PV}}$.

The paper is organized as follows. In Sec. II, at first we take the $\gamma pp$ and $Zpp$ interactions in the low energy limit as examples to  write down the $\gamma Z$-exchange amplitudes and express the amplitudes as linear sums of some general invariant amplitudes. Then we discuss our approach to calculate the corresponding coefficients of the invariant amplitudes. The relations between the invariant amplitudes and the helicity amplitudes in the center-of-mass frame are also given. In Sec. III, we give the analytic expressions for the $\gamma Z$-exchange contributions to the coefficients and helicity amplitudes in the low energy limit. For comparison, the corresponding contributions to the physical quantity $A_{\textrm{PV}}$ are also given. In Sec. IV, we present the numeric comparison between the analytic results and the full numeric results. In Sec. V, we discuss the power behavior of these contributions and explore certain properties of the results when other interactions are considered as inputs. Finally, in Secs. VI and VII, we apply the obtained results to the upcoming P2 experiment and provide a concise summary, respectively.

\section{Basic Formulas}
\subsection{$\gamma Z$-exchange contributions in $ep\rightarrow ep$ at low energy}
\begin{figure}[htbp]
\centering
\includegraphics[height=4.5cm]{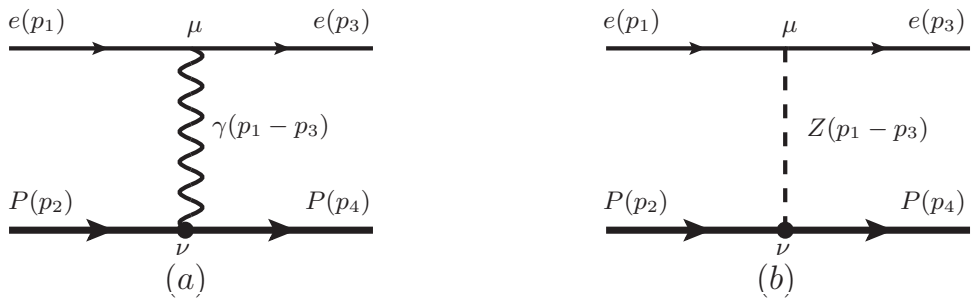}
\caption{The tree diagrams for $ep\rightarrow ep$: (a) represents the one-photon-exchange diagram and (b) represents the one-$Z$-exchange diagram.}
\label{Figure-ep-ep-OBE}
\end{figure}

For the elastic $ep$ scattering, the parity-conserving diagram in the LO of the coupling constant is shown as Fig.~\ref{Figure-ep-ep-OBE}(a) where we label the momenta of the incoming electron, the incoming proton, the outgoing electron, and the outgoing proton as $p_{1,2,3,4}$, respectively. The parity-violating contribution in the LO of coupling constant is from the one-$Z$-exchange diagram, as shown in  Fig.~\ref{Figure-ep-ep-OBE}(b).

When one goes beyond the tree level, the radiative corrections should be considered. Among all the virtual radiative corrections, the contributions from $\gamma Z$ exchange and $WW$ exchange play special roles since their contributions are not only dependent on the momentum transfer but also dependent on the center-of-mass energy. The contributions from $WW$ exchange \cite{ep-ep-gammaZ-zero-energy} can be well estimated since their contributions are dominated in the region where the two $W$ bosons' momenta are large, while the contributions from $\gamma Z$ exchange are much different. In this work, we limit our discussion in the low energy limit where the momentum transfer goes to zero and the center-of-mass energy goes to the minimum physical value at fixed momentum transfer. In this limit, a naive picture is that only the interactions with the LO and the NLO of the momenta give the main contributions. This argument has been used to estimate the TPE contributions in $ep,\mu p$ scattering \cite{lp-lp-TPE-chiral}. In this work, we take the similar assumption to discuss the low energy behaviors of the $\gamma Z$-exchange contributions  at first, and then go back check their validity.

\begin{figure}[htbp]
\centering
\includegraphics[height=8.5cm]{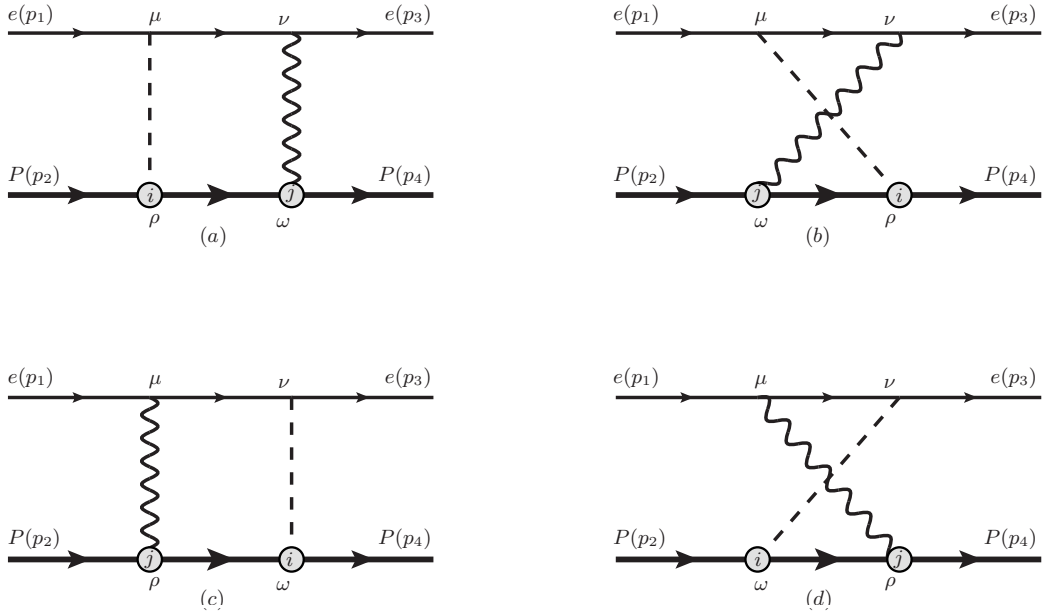}
\caption{The $\gamma Z$-exchange diagrams of $ep\rightarrow ep$: (a,c)  represent the box diagrams and (b,d) represent the crossed-box diagrams.}
\label{Figure-general-gammaZ-exchange-in-ep}
\end{figure}

Naively, for the $\gamma Z$-exchange contributions  in the low energy limit, only the elastic intermediate state gives the contributions, which can be described by the diagrams shown in Fig. \ref{Figure-general-gammaZ-exchange-in-ep}.
Generally, the interactions between the vector bosons and the proton depend on the momentum transfer. In the literature, form factors are commonly introduced to describe the structure of the proton. At low energy scales, the form factors can be expanded order by order in terms of the momentum transfer. The LO and the NLO interactions can be expressed as follows:
\begin{eqnarray}
\Gamma^{\mu}_{\gamma ee} &=&  -ie\gamma^{\mu}, ~~~~\Gamma^{\mu}_{Zee} = -i[\bar{g}_{e}^{V}\gamma^{\mu}+\bar{g}_{e}^{A}\gamma^{\mu}\gamma_{5}],\nonumber \\
\Gamma^{\mu}_{\gamma pp,0} &=&  ieF_1\gamma^{\mu},~~~~\Gamma^{\mu}_{\gamma pp,1}=  ieF_2\frac{i\sigma^{\mu\nu}}{2M_{N}}q_{\nu}, \nonumber \\
\Gamma^{\mu}_{Zpp,0} &=& -i[\bar{g}_1\gamma^{\mu}+\bar{g}_3\gamma^{\mu}\gamma^{5}], ~~~~\Gamma^{\mu}_{Zpp,1} = -i\bar{g}_2\frac{i\sigma^{\mu\nu}}{2M_{N}}q_{\nu}
\label{eq-vertex}
\end{eqnarray}
with
\begin{eqnarray}
\bar{g}_{e}^{V,A}&=&-\frac{e}{4\sin\theta_{\textrm{w}}\cos\theta_{\textrm{w}}}g_{e}^{V,A},\nonumber \\
\bar{g}_{1,2,3}&=&-\frac{e}{4\sin\theta_{\textrm{w}}\cos\theta_{\textrm{w}}}g_{1,2,3},
\end{eqnarray}
where $\theta_{\textrm{w}}$ is the Weinberg angle, $g_{e}^{V}=-1$, $g_{e}^{A}=1-4\sin^2\theta_{\textrm{w}}$ are the coupling constants in the standard model, $F_{1,2}$ are the coupling constants of the $\gamma pp$ interactions in the low energy limit, $g_{1,2,3}$ are the normalized coupling constants of $Zpp$ interactions in the low energy limit, $q$ is the momentum of the incoming boson, the label $0$ refers to the LO interactions which are not dependent on the boson's momentum, and the label 1 refers to the NLO interactions which are proportion to the boson's momentum.

Using these interactions, the corresponding amplitudes for $\gamma Z$-exchange diagrams can be expressed as follows:
\begin{eqnarray}
{\cal M}^{(a)}_{ij} &=& -i\bar{\mu}^{2\epsilon} \int\frac{\textrm{d}^dl}{(2\pi)^d}\bar{u}_{3} \Gamma^{\nu}_{\gamma ee} S_e(p_3+l) \Gamma^{\mu}_{Zee} u_{1} \bar{u}_{4} \Gamma^{\omega}_{\gamma pp,j}(p_4,p_4-l) S_p(p_4-l) \Gamma^{\rho}_{Zpp,i}(p_4-l,p_2) u_{2} \nonumber\\
~~~~~~&& \times D_{\nu\omega}^{\gamma}(l) D_{\mu\rho}^{Z}(p_1-p_3-l), \nonumber \\
{\cal M}^{(b)}_{ij} &=& -i\bar{\mu}^{2\epsilon} \int\frac{\textrm{d}^dl}{(2\pi)^d}\bar{u}_{3} \Gamma^{\nu}_{\gamma ee} S_e(p_3+l) \Gamma^{\mu}_{Zee} u_{1} \bar{u}_{4} \Gamma^{\rho}_{Zpp,i}(p_4,p_2+l) S_p(p_2+l) \Gamma^{\omega}_{\gamma pp,j}(p_2+l,p_2) u_{2} \nonumber\\
~~~~~~&& \times D_{\nu\omega}^{\gamma}(l) D_{\mu\rho}^{Z}(p_1-p_3-l), \nonumber \\
{\cal M}^{(c)}_{ij} &=& -i\bar{\mu}^{2\epsilon} \int\frac{\textrm{d}^dl}{(2\pi)^d}\bar{u}_{3} \Gamma^{\nu}_{Zee} S_e(p_1-l) \Gamma^{\mu}_{\gamma ee} u_{1} \bar{u}_{4} \Gamma^{\omega}_{Zpp,i}(p_4,p_2+l) S_p(p_2+l) \Gamma^{\rho}_{\gamma pp,j}(p_2+l,p_2) u_{2} \nonumber\\
~~~~~~&& \times D_{\nu\omega}^{\gamma}(l) D_{\mu\rho}^{Z}(p_1-p_3-l), \nonumber \\
{\cal M}^{(d)}_{ij} &=& -i\bar{\mu}^{2\epsilon} \int\frac{d^dl}{(2\pi)^d}\bar{u}_{3} \Gamma^{\nu}_{Zee} S_e(p_1-l) \Gamma^{\mu}_{\gamma ee} u_{1} \bar{u}_{4} \Gamma^{\rho}_{\gamma pp,j}(p_4,p_4-l) S_p(p_4-l) \Gamma^{\omega}_{Zpp,i}(p_4-l,p_2) u_{2} \nonumber\\
~~~~~~&& \times D_{\nu\omega}^{\gamma}(l) D_{\mu\rho}^{Z}(p_1-p_3-l),
\label{eq-amplitudes-2gamma}
\end{eqnarray}
where $\bar{u}_{i},u_{i}$ are the short writing of the spinors $\bar{u}(p_i,s_i),u(p_i,s_i)$ with corresponding masse $m_i$, respectively, $l$ is the momentum of the photon, $i$ and $j$ refer to the order of the momentum in the vertices $\Gamma_{\gamma pp}$ and $\Gamma_{Zpp}$, $\bar{\mu}$ is the renormalization scale and $\epsilon=4-d$ with $d$ the dimension. In the naive picture, the label $ij=00$ corresponds to the LO contribution, while $ij=01$ and $ij=10$ correspond. Additionally, $ij=22$ corresponds to the next-to-next-leading-order (NNLO) contribution. In practical calculations, we consider all of these contributions for comparison.

In practical calculations, the package FeynCalc10.0 \cite{FenyCalc} is used to deal with Dirac matrix, the PackageX3.0 \cite{PacakgeX} is used to do the loop integration and the the package LoopTools \cite{LoopTools} is used for cross-check.

\subsection{The scheme to handle $\gamma_5$ in $d$ dimension and the general invariant amplitudes}
To calculate the amplitudes ${\cal M}^{(a,b,c,d)}_{ij}$ using the dimension regularization, one must select a scheme to handle the Dirac matrix $\gamma_5$ in $d$ dimension. This differs slightly from similar calculation in the parity-conserving $ep$ scattering, where, in principle, there is no $\gamma_5$ at the amplitude level in the latter case. In our practical calculation, we choose the NDR scheme in FeynCalc~\cite{FenyCalc} to handle $\gamma_5$.

In the NDR scheme, there is an ambiguous definition for the trace of a matrix with an odd number of $\gamma_5$ matrices. To avoid this ambiguity in the calculation, we separate the full amplitude into a parity conserved (PC) part and a parity violated (PV) part. We then further separate the PV part as follows:
\begin{eqnarray}
{\cal M}^{(a+b+c+d)}_{\gamma Z} &\equiv& {\cal M}^{\textrm{PC}}_{\gamma Z}+{\cal M}^{\textrm{\textrm{PV}}}_{\gamma Z},\nonumber\\
{\cal M}^{\textrm{\textrm{PV}}}_{\gamma Z} &\equiv& \bar{g}_{e}^A {\cal M}^{V}_{\gamma Z}+ \bar{g}_{e}^V {\cal M}^{A}_{\gamma Z}.
\label{eq-separate-V-A}
\end{eqnarray}
Since our focus is on the PV part of the amplitude, we will only discuss the amplitudes ${\cal M}^{V}_{\gamma Z}$ and ${\cal M}^{A}_{\gamma Z}$ in the following. After making the approximation of $m_e = 0$, where $m_e$ represents the mass of electron,  the amplitudes ${\cal M}^{V,A}_{\gamma Z}$ can be expressed as follows:
\begin{eqnarray}
{\cal M}^{V}_{\gamma Z} &\equiv& \sum_{i=1}^{3} {\cal F}_{\gamma Z,i}^{V}{\cal P}_{i}^{V}\equiv\sum \limits_{i=1}^{3}\sum \limits_{j=1}^{2}\sum \limits_{k=1}^{2}{\cal C}_{\gamma Z,ijk}^{V}F_j \bar{g}_k{\cal P}_{i}^{V},\nonumber\\
{\cal M}^{A}_{\gamma Z}&\equiv& \sum_{i=1}^{3} {\cal F}_{\gamma Z,i}^{A}\mathcal{P}_{i}^{A}
\equiv\sum \limits_{i=1}^{3}\sum \limits_{j=1}^{2}{\cal C}_{\gamma Z,ij3}^{A}F_j \bar{g}_3{\cal P}_{i}^{A},
\label{eq-amplitude-total}
\end{eqnarray}
where the general invariant amplitudes ${\cal P}_{i}^{V}$ and ${\cal P}_{i}^A$ are chosen as
\begin{eqnarray}
{\cal P}_{1}^{V} &\equiv& \left[\bar{u}_{3} \gamma_{\mu} \gamma_{5} u_{1}\right]\left[\bar{u}_{4} \gamma^{\mu} u_{2}\right], \nonumber\\
{\cal P}_{2}^{V} &\equiv& \frac{1}{Q}\left[\bar{u}_{3} \gamma_{\mu} \gamma_{5} u_{1}\right]\left[\bar{u}_{4} i\sigma^{\mu\nu}q_\nu u_{2}\right],  \nonumber \\
{\cal P}_{3}^{V} &\equiv& \frac{1}{M_NQ}\left[\bar{u}_{3}\sla{P}\gamma_{5} u_{1}\right]\left[\bar{u}_{4}\sla{K}u_{2}\right],\nonumber \\
{\cal P}_{1}^{A} &\equiv& \left[\bar{u}_{3} \gamma^{\mu}  u_{1}\right]\left[\bar{u}_{4} \gamma_{\mu}\gamma_{5} u_{2}\right],  \nonumber \\
{\cal P}_{2}^{A} &\equiv& \frac{1}{Q}\left[\bar{u}_{3} \gamma^{\mu} u_{1}\right]\left[\bar{u}_{4} \gamma_{\mu} \sla{K} \gamma_{5} u_{2}\right],  \nonumber \\
{\cal P}_{3}^{A} &\equiv& \frac{1}{M_NQ}\left[\bar{u}_{3}\sla{P} u_{1}\right]\left[\bar{u}_{4} \sla{K}  \gamma_{5}u_{2}\right]
\label{eq-invariant-amplitudes-reference}
\end{eqnarray}
with $P=p_2+p_4$, $K=p_1+p_3$, $Q^2=-q^2$, $q=p_4-p_2=p_1-p_3$, and $M_N$ the mass of proton.

This separation differs slightly from the form used in the references, where typically only three invariant amplitudes are chosen \cite{ep-ep-gammaZ-dispersion-relation-1}.  As we have argued earlier,  the purpose of this separation is to avoid the ambiguous definition of $\gamma_5$ in $d$ dimensions. With these definitions,  the calculation of the coefficients $C_{\gamma Z,ijk}^{V}$ and $C_{\gamma Z,ij3}^{A}$ now involves only even powers of $\gamma_5$.

Similarly, we separate the amplitude for one-$Z$ exchange, ${\cal M}_Z$, as
\begin{eqnarray}
{\cal M}_{ Z} &\equiv& {\cal M}^{\textrm{PC}}_{  Z}+{\cal M}^{\textrm{\textrm{PV}}}_{Z},\nonumber\\
{\cal M}^{\textrm{\textrm{PV}}}_{Z} &\equiv& \bar{g}_{e}^A {\cal M}^{V}_{Z}+ \bar{g}_{e}^V {\cal M}^{A}_{Z},\nonumber\\
{\cal M}_{Z}^{X}&\equiv&\sum_{i=1}^{3}{\cal F}_{Z,i}^{X}{\cal P}_{i}^{X}
\end{eqnarray}
with $X=V$ or $A$, respectively.

\subsection{Calculation of $C_{\gamma Z,ijk}^{V}$ and $C_{\gamma Z,ijk}^{A}$}

To calculate the coefficients $C_{\gamma Z,ijk}^{V}$ and $C_{\gamma Z,ijk}^{A}$, one can solve the following system of algebraic equations in $d$ dimensions:
\begin{eqnarray}
\sum_{helicity}{\cal M}^{V}_{\gamma Z}{\cal T}_{n}^{V*} &=& \sum_{helicity}\sum_{i=1}^{3} {\cal F}_{\gamma Z,i}^{V}{\cal P}_{i}^{V}{\cal T}_{n}^{V*}=\sum_{helicity}\sum \limits_{i=1}^{3}\sum \limits_{j=1}^{2}\sum \limits_{k=1}^{2}{\cal C}_{\gamma Z,ijk}^{V}F_j \bar{g}_k{\cal M}_{i}^{V}{\cal T}_{n}^{V*},\nonumber\\
\sum_{helicity}{\cal M}^{A}_{\gamma Z}{\cal T}_{n}^{V*}&=& \sum_{helicity}\sum_{i=1}^{3} {\cal F}_{\gamma Z,i}^{A}{\cal P}_{i}^{A}{\cal T}_{n}^{A*} = \sum_{helicity}\sum \limits_{i=1}^{3}\sum \limits_{j=1}^{2}{\cal C}_{\gamma Z,ij3}^{A}F_j \bar{g}_3{\cal M}_{i}^{A}{\cal T}_{n}^{A*},
\label{eq-algebraic-equations}
\end{eqnarray}
where ${\cal T}_{n}^{V}$ and ${\cal T}_{n}^{A}$ can be directly chosen as ${\cal P}_{n}^{V}$ and ${\cal P}_{n}^{A}$, respectively.

After calculating the following matrix in $d$-dimension:
\begin{eqnarray}
\mathcal{D}_{ij}^{X} &\equiv & \sum_{helicity} {\cal P}_{i}^{X} {\cal P}_{j}^{X*},
\end{eqnarray}
the coefficients ${\cal F}^{X}_{\gamma Z,i}$ can be expressed as
\begin{eqnarray}
{\cal F}^{X}_{\gamma Z,i} &=& \sum_{j}[(\mathcal{D}^{X})^{-1}]_{ij}\sum_{helicity}{\cal M}^{X}_{\gamma Z}{\cal P}_{j}^{*}.
\label{eq-coefficient}
\end{eqnarray}
Once ${\cal F}^{X}{\gamma Z,i}$ is known, the corresponding coefficients $C{\gamma Z,ijk}^{X}$ can be directly obtained.

The expressions of $\mathcal{D}_{ij}^{X}$ in $d$ dimensions are a little complex, so we do not list them here.

\subsection{From general invariant amplitudes to helicity amplitudes}
In some cases, the physical meaning of the general invariant amplitudes and their coefficients may not be clear, as the behaviors of the coefficients $C_{\gamma Z,ijk}^{X}$ can include certain kinematic effects. Conversely, the physical meaning of the helicity amplitudes is much clearer.

After performing some simple calculations, one can observe the following properties in the center-of-mass frame:
\begin{eqnarray}
&&{\cal M}_{Y}^{+-\pm\pm,\textrm{\textrm{PV}}}={\cal M}_{Y}^{-+\pm\pm,\textrm{\textrm{PV}}}=0,\nonumber\\
&&{\cal M}_{Y}^{++++,\textrm{\textrm{PV}},} = -{\cal M}_{Y}^{----,\textrm{\textrm{PV}}},\nonumber\\
&&{\cal M}_{Y}^{+++-,\textrm{\textrm{PV}}} =- {\cal M}_{Y}^{++-+,\textrm{\textrm{PV}}} =- {\cal M}_{Y}^{--+-,\textrm{\textrm{PV}}} = {\cal M}_{Y}^{---+,\textrm{\textrm{PV}}},\nonumber\\
&&{\cal M}_{Y}^{++--,\textrm{\textrm{PV}}} =-{\cal M}_{Y}^{--++,\textrm{\textrm{PV}}},
\end{eqnarray}
where the index $Y$ refers to either $Z$ or $\gamma Z$, and the indexes such as ++++ correspond to the helicities of the incoming electron, the outgoing electron, the incoming proton, and the outgoing proton, respectively.

The helicity amplitudes can be expressed as follows:
\begin{eqnarray}
{\cal M}_{Y}^{\pm\pm\pm\pm,X} &=& \sum_{i} {\cal F }^{X}_{Y,i}{\cal P}_i^{\pm\pm\pm\pm,X}.
\label{Eq:invariant-amplitude-helicity-amplitude}
\end{eqnarray}
In Table \ref{Tab-helicity-amplitude-of-invariant-amplitude}, we present the expressions for ${\cal P}_i^{\pm\pm\pm\pm,X}$ in the center-of-mass frame, where the momenta are chosen as
\begin{eqnarray}
p_1^{\mu}&=&(E_c,0,0,E_c),\nonumber\\
p_2^{\mu}&=&(\sqrt{M_N^2+E_c^2},0,0,-E_c), \nonumber\\
p_3^{\mu}&=&(E_c,E_c\sin\theta_c,0,E_c\cos\theta_c),
\end{eqnarray}
and some variables are defined as
\begin{eqnarray}
s &\equiv& (p_1+p_2)^2,\nonumber\\
\nu &\equiv& 2s-2M_{N}^2-Q^2, \nonumber \\
a&\equiv&\nu+Q^2+2M_NQ,\nonumber\\
b&\equiv&\nu+Q^2-2M_NQ,\nonumber\\
c&\equiv&\sqrt{\nu^2-4M_N^2Q^2-Q^4}.
\end{eqnarray}

\begin{table}[htbp]
\centering
\begin{tabular}{p{2.5cm}<{\centering}p{2.5cm}<{\centering}p{2.5cm}<{\centering}p{2.5cm}<{\centering}}
\hline
$i$ & $1$ & $2$ & $3$  \\
\hline
$ {\cal P}^{++++,V}_i$ & $-\frac{ab}{\nu+Q^2}$ & $-4M_NQ$ & $-\frac{c^2}{M_NQ}$ \\
\hline
$ {\cal P}^{+++-,V}_i$ & $\frac{2cM_NQ}{\nu+Q^2}$ & $-c$ & $2c$ \\
\hline
$ {\cal P}^{++--,V}_i$ & $-\frac{c^2}{\nu+Q^2}$ & $0$ & $-\frac{c^2}{M_NQ}$ \\
\hline
\hline
$ {\cal P}^{++++,A}_i$ & $-\frac{ab+8M_N^2Q^2}{\nu+Q^2}$ & $0$ & $-\frac{c^2}{M_NQ}$ \\
\hline
$ {\cal M}^{+++-,A}_i$ & $-\frac{2cM_NQ}{\nu+Q^2}$ & $-c$ & 0 \\
\hline
$ {\cal M}^{++--,A}_i$ & $\frac{c^2}{\nu+Q^2}$ & $0$ & $\frac{c^2}{M_NQ}$ \\
\hline
\end{tabular}
\caption{The expressions for the invariant amplitudes with special helicities ${\cal P}_i^{\pm\pm\pm\pm,X}$ in the center-of-mass frame.}
\label{Tab-helicity-amplitude-of-invariant-amplitude}
\end{table}

Using these expressions, the helicity amplitudes ${\cal M}_{\gamma Z}^{\pm\pm\pm\pm,X}$ can be expressed as direct linear combinations of the coefficients $C_{\gamma Z,ijk}^{X}$. For instance, we have the following relationship:
\begin{eqnarray}
{\cal M}^{++++,V}_{\gamma Z} &=& -\frac{ab}{\nu+Q^2}{\cal F}_{\gamma Z,1}^{V}-4M_NQ{\cal F}_{\gamma Z,2}^{V}-\frac{c^2}{M_NQ}{\cal F}_{\gamma Z,3}^{V} \nonumber\\
&=&-\sum \limits_{j=1}^{2}\sum \limits_{k=1}^{2}\Big[\frac{ab}{\nu+Q^2}C_{\gamma Z,1jk}^{V}+4M_NQC_{\gamma Z,2jk}^{V}+\frac{c^2}{M_NQ}C_{\gamma Z,3jk}^{V}\Big]F_j\bar{g}_k.
\end{eqnarray}

\section{Expressions in the Low Energy limit }

In this section, we firstly present the analytical expressions for $C_{\gamma Z,ijk}^{V,A}$ in the low-energy limit. Then, we provide the analytical expressions for the corresponding corrections to the helicity amplitudes and the experimental measurement $A_{\textrm{PV}}$. These analytic expressions clearly demonstrate the power-law behavior of the corrections in a transparent manner. Furthermore, they can be utilized to estimate the $\gamma Z$-exchange corrections for all feasible future measurements, particularly at low values of $Q$ and $E$, assuming knowledge of the low-energy constants.

\subsection{$C_{\gamma Z,ijk}^{X}$ when $Q^2\rightarrow 0$ and $\nu\rightarrow \nu_{min}$ }
Before discussing the properties of the coefficients $C_{\gamma Z,ijk}^{X}$ in the low energy limit, for comparison, we list the expressions of ${\cal F}_{Z,i}^{X}$ as
\begin{eqnarray}
\textrm{Re}[{\cal F}_{Z,1}^{V}]&=& -\frac{\bar{g}_1}{M_Z^2},
~~~~\textrm{Re}[{\cal F}_{Z,2}^{V}]= -\frac{\bar{g}_2}{M_Z^2}\frac{Q}{2M_N},
~~~~\textrm{Re}[{\cal F}_{Z,3}^{V}]= 0, \nonumber\\
\textrm{Re}[{\cal F}_{Z,1}^{A}]&=&-\frac{\bar{g}_3}{M_Z^2},
~~~~\textrm{Re}[{\cal F}_{Z,2}^{A}]= 0,
~~~~~~~~~~~~~~~~~~\textrm{Re}[{\cal F}_{Z,3}^{A}]= 0.
\end{eqnarray}

Physically, when $Q^2$ is fixed, the physical $\nu$ has a minimum value given by
\begin{eqnarray}
\nu_{phs} \geq \nu_{min}=Q\sqrt{4M_N^2+Q^2}.
\label{eq-minnimum-of-v}
\end{eqnarray}
To calculate $C_{\gamma Z,ijk}^{X}$ in the low energy limit, we expand them at $Q\rightarrow0$ and $\delta\equiv\nu-\nu_{min}\rightarrow 0$ independently after the loop integration.  This approach differs slightly from the usual discussion where $Q\rightarrow 0$ and $E_e\rightarrow 0$ are used, with $E_e$ the energy of initial electron in the laboratory frame. The reason for not expanding the results in terms of $Q$ and $E_e$ is that when $Q$ is fixed, $E_e$ actually has a minimum value given by
\begin{eqnarray}
E_e \geq E_{min}=\frac{\nu_{min}+Q^2}{4M_N}.
\label{eq-minnimum-of-v}
\end{eqnarray}
This implies  that $Q\rightarrow 0$ and $E_e\rightarrow 0$ are not completely independent for the physical process.

In practical calculations, the expansion of  $C_{\gamma Z,ijk}^{X}$ on $Q$ and $\delta$ should be done independently. This means that the expansion is valid for any $\delta/Q$ ratio. The final expressions for the nonzero LO contributions $C_{\gamma Z,ijk}^{X,\textrm{LO}}$ are presented in the Appendix A.

We would like to mention that the contributions which are only dependent on $Q$ but not dependent on $\delta$, have the similar behaviors with the radiative corrections to the vertexes $\Gamma^{\mu}_{\gamma pp}$ and $\Gamma^{\mu}_{Zpp}$. This means that these contributions can be absorbed into certain constants at fixed $Q$. However, the most significant characteristic of the $\gamma Z$-exchange contributions lies in their dependence on $\delta$ at a fixed $Q$.

\subsection{Helicity amplitudes when $Q^2\rightarrow 0$ and $\nu\rightarrow \nu_{min}$}

Since the physical meaning of the helicity amplitudes is much more definite than that of the coefficients, we provide the expressions for the helicity amplitudes in the low-energy limit in this subsection.

At the tree level, the helicity amplitudes due to the one-$Z$-exchange can be separated as
\begin{eqnarray}
{\cal M}_{Z}^{\pm\pm\pm\pm,V}&\equiv& \sum_{k=1}^{2} {\cal M}_{Z,k}^{\pm\pm\pm\pm,V}\bar{g}_k\bar{g}_e^A, \nonumber\\
{\cal M}_{Z}^{\pm\pm\pm\pm,A}&\equiv& {\cal M}_{Z,3}^{\pm\pm\pm\pm,A}\bar{g}_3\bar{g}_e^V.
\end{eqnarray}
After expanding ${\cal M}^{\pm\pm\pm\pm,X}_{Z,k}$ on $Q$ and $\delta$ independently, the non-zero LO contributions ${\cal M}^{\pm\pm\pm\pm,X,\textrm{LO}}_{Z,k}$ are expressed as Table \ref{Tab-helicity-amplitude-one-Z-exchange} with
\begin{eqnarray}
h&=&\frac{1}{M_Z^2(2M_NQ+\delta)},\nonumber\\
z&=&\sqrt{(4M_NQ+\delta)\delta}.
\end{eqnarray}

\begin{table}[htbp]
\centering
\begin{tabular}{p{2.5cm}<{\centering}p{3.2cm}<{\centering}p{3.2cm}<{\centering}p{3.2cm}<{\centering}}
\hline
$\textrm{Hs}$ & $++++$ & $+++-$ & $++--$  \\
\hline
${\cal M}^{\textrm{Hs},V,\textrm{LO}}_{Z,1}$ & $hz^2$ & $-2hzM_NQ$ & $hz^2$ \\
\hline
${\cal M}^{\textrm{Hs},V,\textrm{LO}}_{Z,2}$ & $2h(2M_NQ+\delta)Q^2$ & $-hz \frac{(2M_NQ+\delta)Q}{2M_N}$ & $0$ \\
\hline
${\cal M}^{\textrm{Hs},A,\textrm{LO}}_{Z,3}$ & $h(8M_N^2Q^2+z^2)$ & $2hz M_NQ$ & $-hz^2$ \\
\hline
\end{tabular}
\caption{The expressions for ${\cal M}^{\pm\pm\pm\pm,X,\textrm{LO}}_{Z,k}$ in the center-of-mass frame.}
\label{Tab-helicity-amplitude-one-Z-exchange}
\end{table}
Similarly, the helicity amplitudes due to the $\gamma Z$ exchange can be separated  as
\begin{eqnarray}
{\cal M}_{\gamma Z}^{\pm\pm\pm\pm,V}&\equiv& \sum_{jk} {\cal M}_{\gamma Z,jk}^{\pm\pm\pm\pm,V} F_j\bar{g}_k\bar{g}_e^A, \nonumber\\
{\cal M}_{\gamma Z}^{\pm\pm\pm\pm,A}&\equiv& \sum_{j} {\cal M}_{\gamma Z,j3}^{\pm\pm\pm\pm,A}F_j\bar{g}_3\bar{g}_e^V.
\end{eqnarray}

We would like to mention that in order to obtain the correct expressions for the nonzero LO contributions ${\cal M}_{\gamma Z,jk}^{\pm\pm\pm\pm,X,\textrm{LO}}$, one should not directly substitute $C_{\gamma Z,ijk}^{X,\textrm{LO}}$ into Eq. (\ref{Eq:invariant-amplitude-helicity-amplitude}). The reason for this is that there are cancellations between the contributions from different $C_{\gamma Z,ijk}^{X,\textrm{LO}}$ at the LO in certain specific cases. Therefor, to obtain the correct expressions for the nonzero LO contributions ${\cal M}_{\gamma Z,jk}^{\pm\pm\pm\pm,X,\textrm{LO}}$, one should substitute $C_{\gamma Z,ijk}^{X}$ into Eq. (\ref{Eq:invariant-amplitude-helicity-amplitude}), and then expand ${\cal M}_{\gamma Z,jk}^{\pm\pm\pm\pm,X}$ as $Q\rightarrow0$ and $\delta\rightarrow 0$.  The existence of these cancellations also suggests that the helicity amplitudes reflect the physical properties in a more definite manner.

The final expressions for the nonzero LO contributions ${\cal M}_{\gamma Z,jk}^{\pm\pm\pm\pm,X,\textrm{LO}}$ are presented in Appendix B.

\subsection{$A_{\textrm{PV}}$ when $Q^2\rightarrow 0$ and $\nu\rightarrow \nu_{min}$ }

After obtaining the coefficients $C_{\gamma Z,ijk}^{X}$ or ${\cal M}{\gamma Z, jk}^{\pm\pm\pm\pm,X}$, the $\gamma Z$-exchange contributions to all related physical quantities can be determined. In this subsection, we discuss the $\gamma Z$-exchange contributions to the physical measurement $A_{\textrm{PV}}$  which is defined as
\begin{eqnarray}
A_{\textrm{PV}} &\equiv&\frac{\sum\limits_{helicity}({\cal M}^{+}{\cal M}^{+*}-{\cal M}^{-}{\cal M}^{-*})}{\sum\limits_{helicity}({\cal M}^{+}{\cal M}^{+*}+{\cal M}^{-}{\cal M}^{-*})}, \end{eqnarray}
where ${\cal M}^{+,-}$ are the helicity amplitudes in the laboratory frame with the helicity of the incoming electron being $\pm$, respectively.

At the tree level, where only the contributions from the one-photon exchange and the one-$Z$ exchange diagrams are considered, the corresponding $A_{\textrm{PV}}^{\gamma\otimes Z}$ can be expressed as
\begin{eqnarray}
A_{\textrm{PV}}^{\gamma\otimes Z}&=&\frac{1}{e^2\sigma}\Big[\sum_{i=1}^{2}\sum_{k=1}^{2}{\cal A}_{Z,ik}^{V}F_i\bar{g}_{k}\bar{g}_e^{A}+\sum_{i=1}^{2}{\cal A}_{Z,i3}^{A}F_i\bar{g}_3\bar{g}_e^{V}\Big],
\end{eqnarray}
where
\begin{eqnarray}
\sigma &=&4F_1^2M_N^2(\nu^2-4M_N^2Q^2+Q^4)+F_2^2Q^2(\nu^2+4M_N^2Q^2-Q^4)+16F_1F_2M_N^2Q^4\nonumber\\
&\rightarrow&4F_1^2M_N^2z^2
\end{eqnarray}
and
\begin{eqnarray}
{\cal A}_{Z,11}^{V}&=&-\frac{8}{M_Z^2}M_N^2Q^2(\nu^2-4M_N^2Q^2+Q^4)\rightarrow-\frac{4}{M_Z^2}M_NQ^2(4M_NQ+\delta)(2M_N\delta+Q^3), \nonumber\\
{\cal A}_{Z,12}^{V}&=&-\frac{16}{M_Z^2}M_N^2Q^6, \nonumber\\
{\cal A}_{Z,21}^{V}&=&-\frac{16}{M_Z^2}M_N^2Q^6,\nonumber\\
{\cal A}_{Z,22}^{V}&=&-\frac{2}{M_Z^2}Q^4(\nu^2+4M_N^2Q^2-Q^4)\rightarrow-\frac{2}{M_Z^2}Q^4(8M_N^2Q^2+z^2), \nonumber\\
{\cal A}_{Z,13}^{A}&=&-\frac{16}{M_Z^2}M_N^2Q^4\nu\rightarrow-\frac{16}{M_Z^2}M_N^2Q^4(2M_NQ+\delta) , \nonumber\\
{\cal A}_{Z,23}^{A}&=&-\frac{16}{M_Z^2}M_N^2Q^4\nu\rightarrow-\frac{16}{M_Z^2}M_N^2Q^4(2M_NQ+\delta).
\label{eq-expression-of-Hij}
\end{eqnarray}

When considering the interference between the one-photon exchange diagram and $\gamma Z$-exchange diagrams, the corresponding $A_{\textrm{PV}}^{\gamma\otimes\gamma Z}$ is expressed as
\begin{eqnarray}
A_{\textrm{PV}}^{\gamma\otimes\gamma Z}
&=&\frac{1}{e^2\sigma}\Big(\sum_{i=1}^{3}{\cal N}_{i}^{V}\textrm{Re}[{\cal F}_{\gamma Z,i}^{V}]\bar{g}_e^{A} +\sum_{i=1}^{3}{\cal N}_{i}^{A}\textrm{Re}[{\cal F}_{\gamma Z,i}^{A}]\bar{g}_e^{V}\Big)\nonumber \\
&=&\frac{1}{e^2\sigma}\Big(\sum_{i=1}^{3}\sum_{j=1}^{2} \sum_{k=1}^{2}{\cal N}_{i}^{V}\textrm{Re}[C_{\gamma Z,ijk}^{V}]F_{j}^{\gamma Z}\bar{g}_{k}\bar{g}_e^{A}+\sum_{i=1}^{3}\sum_{j=1}^{2}{\cal N}_{i}^{A}\textrm{Re}[C_{\gamma Z,ij3}^{A}]F_{j}^{\gamma Z}\bar{g}_3\bar{g}_e^{V}\Big) \nonumber\\
\end{eqnarray}
with
\begin{eqnarray}
{\cal N}_{1}^{V}&=&8M_N^2Q^2[(\nu^2-4M_N^2Q^2+Q^4)F_1+2Q^4F_2], \nonumber\\
{\cal N}_{2}^{V}&=&4M_NQ[8M_N^2Q^4F_1+Q^2(\nu^2+4M_N^2Q^2-Q^4)F_2],\nonumber\\
{\cal N}_{3}^{V}&=&8M_NQ\nu(\nu^2-4M_N^2Q^2-Q^4)F_1,\nonumber\\
{\cal N}_{1}^{A}&=&16M_N^2Q^4\nu(F_1+F_2),\nonumber\\
{\cal N}_{2}^{A}&=&4M_NQ^3[8M_N^2Q^2F_1+(\nu^2+4M_N^2Q^2-Q^4)F_2],\nonumber\\
{\cal N}_{3}^{A}&=&8M_NQ^3(\nu^2-4M_N^2Q^2-Q^4)(F_1+F_2),
\end{eqnarray}
where we have used the indexes $\gamma$ and $\gamma Z$ in $F_i$ to indicate the source of the coupling constants, although their numerical values are equal. For instance, $F_{1}^{\gamma}$ represents the coupling constant from the one-photon exchange diagram, while $F_{1}^{\gamma Z}$ represents the coupling constant from the $\gamma Z$-exchange diagram.

After substituting $C_{\gamma Z,ijk}^{X}$ into the expression, one can express $A_{\textrm{PV}}^{\gamma\otimes\gamma Z}$ as
\begin{eqnarray}
A_{\textrm{\textrm{PV}}}^{\gamma\otimes\gamma Z}&\equiv& \frac{1}{e^2\sigma}\Big\{\sum_{i=1}^{2}\sum_{j=1}^{2}\sum_{k=1}^{2}\textrm{Re}[{\cal A}_{\gamma Z,ijk}^{V}]F_i^{\gamma}F_j^{\gamma Z}\bar{g}_{k}\bar{g}_e^{A}
+\sum_{i=1}^{2}\sum_{j=1}^{2}\textrm{Re}[{\cal A}_{\gamma Z,ij3}^{A}]F_i^{\gamma}F_j^{\gamma Z}\bar{g}_3\bar{g}_e^{V} \Big\}\nonumber \\
&\equiv& \frac{G_Ft}{4\sqrt{2}\pi\alpha_{e}}[\Box_{\gamma Z}^{A}+\Box_{\gamma Z}^{V}],
\end{eqnarray}
where $G_F=\pi\alpha_e/(\sqrt{2}M_Z^2\sin^2\theta_{\textrm{w}}\cos^2\theta_{\textrm{w}})$ is the Fermi constant, $\alpha_e=e^2/4\pi$ is the fine structure constant, and $t=-Q^2$. Similar to ${\cal M}_{\gamma Z,jk}^{\pm\pm\pm\pm,X}$, to obtain the nonzero LO contributions ${\cal A}_{\gamma Z,ijk}^{X,\textrm{LO}}$, one should substitute $C_{\gamma Z,ijk}^{X}$ into the expressions and then expand them in terms of $Q$ and $\delta$ around 0. The final expressions for the nonzero LO contributions ${\cal A}_{\gamma Z,ijk}^{X,\textrm{LO}}$ are presented in Appendix C.

\section{Numerical Properties }
\begin{figure}[htbp]
\centering
\includegraphics[height=12cm]{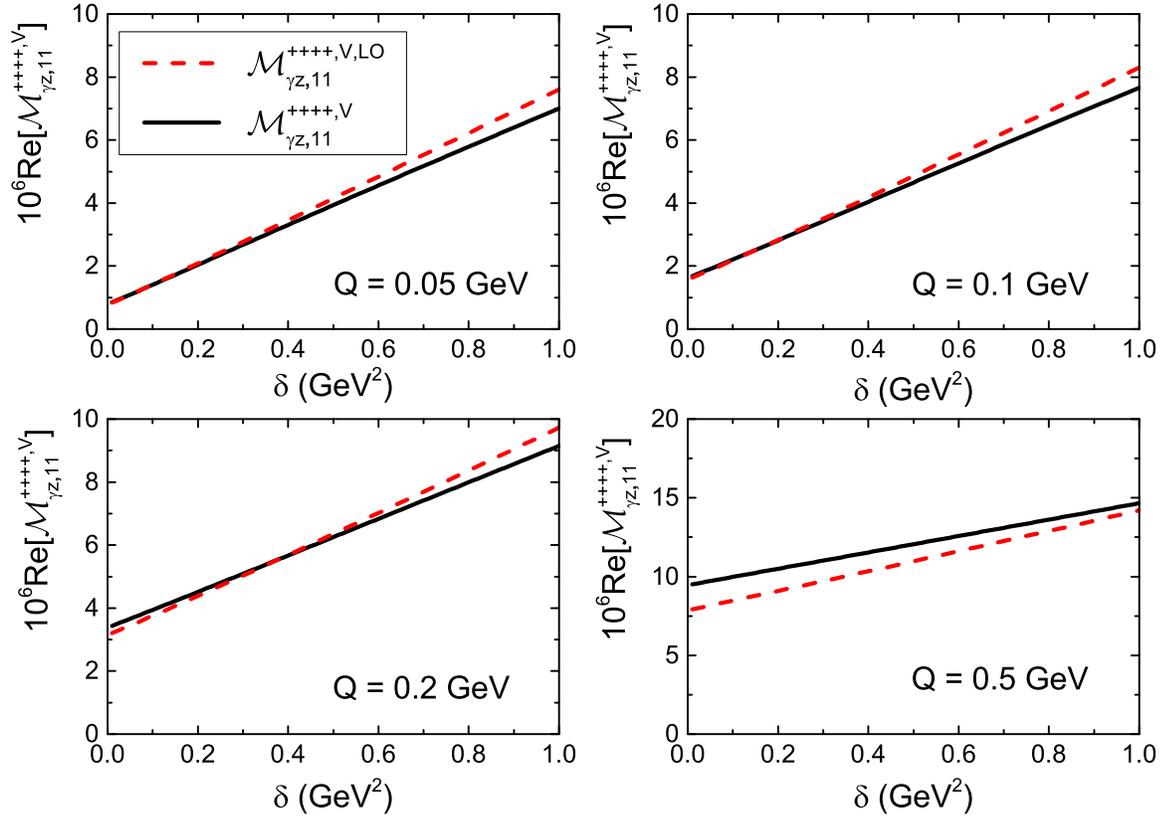}
\caption{Comparison between $\textrm{Re}[{\cal M}^{++++,V,\textrm{LO}}_{\gamma Z,11}]$ and $\textrm{Re}[{\cal M}^{++++,V}_{\gamma Z,11}]$ at $Q=0.05,0.1,0.2,$ and $0.5$ GeV, respectively.}
\label{Figure-Mpp-Re-11}
\end{figure}

\begin{figure}[htbp]
\centering
\includegraphics[height=12cm]{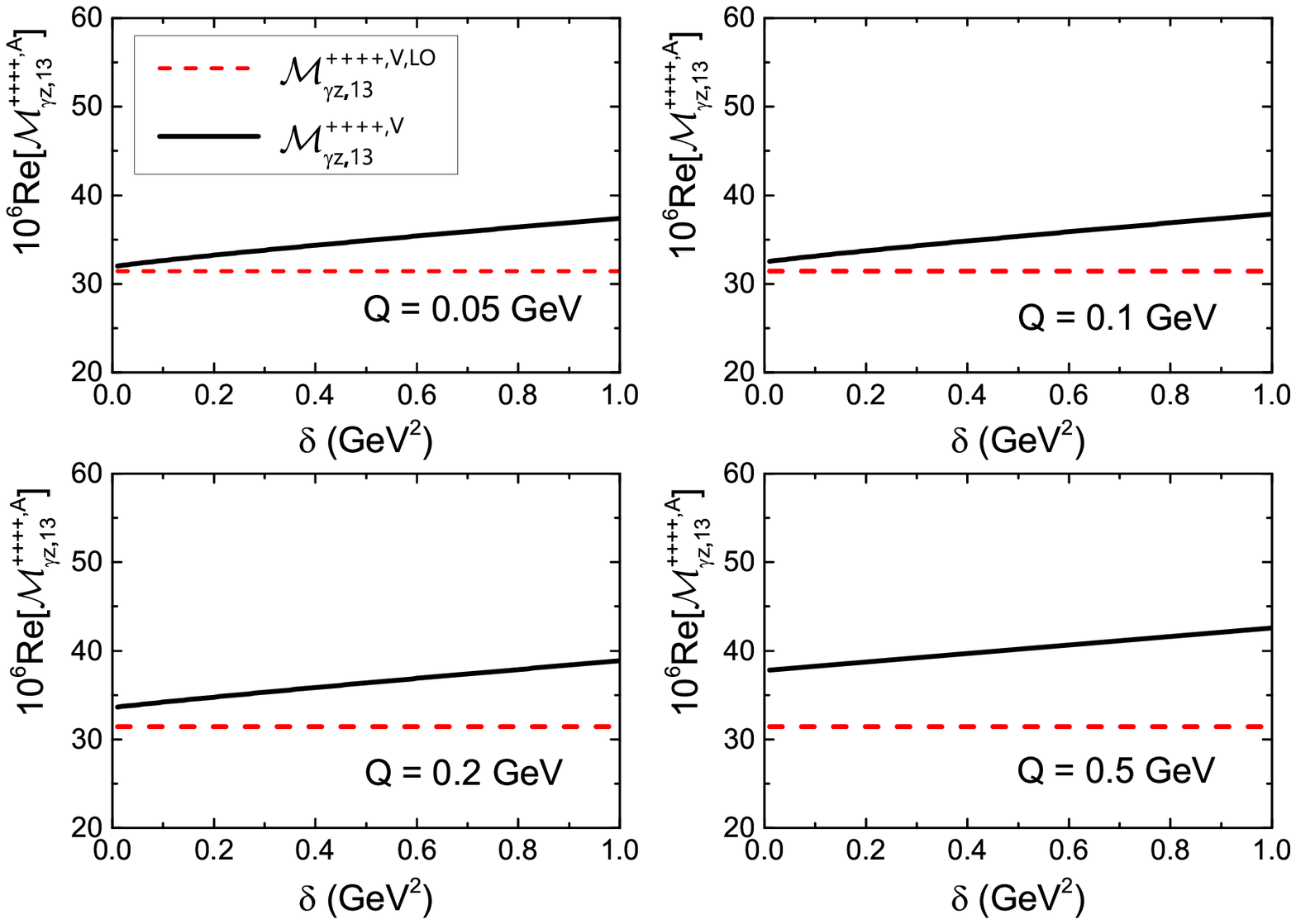}
\caption{Comparison between $\textrm{Re}[{\cal M}^{++++,A,\textrm{LO}}_{\gamma Z,13}]$ and $\textrm{Re}[{\cal M}^{++++,A}_{\gamma Z,13}]$ at $Q=0.05,0.1,0.2,$ and $0.5$ GeV, respectively.}
\label{Figure-Mpp-Re-13}
\end{figure}

Before discussing the analytic properties, we perform a numerical comparison between the non-zero LO contributions and the full contributions for different values of  $Q$ in the range $(0.05, 0.1, 0.2, 0.5)$ GeV and for different values of  $\delta$ in the range $[0, 1]$ GeV$^2$. We find that the two results are very similar across all these regions. In Figs. \ref{Figure-Mpp-Re-11} and \ref{Figure-Mpp-Re-13}, we take $\textrm{Re}[{\cal M}_{\gamma Z,11}^{++++,V,\textrm{LO}}]$ and $\textrm{Re}[{\cal M}_{\gamma Z,13}^{++++,V,\textrm{LO}}]$ as examples to show  the numerical comparison. The comparisons clearly demonstrate that the analytic nonzero LO expressions provide a reliable approximation of the full result at the low energy scale.

\section{Power Behaviors in the Low Energy Limit}
\subsection{Power behaviors of $C_{\gamma Z,ijk}^{X}$ in the low energy limit}

The analytic expressions for $C_{\gamma Z,ijk}^{X,\textrm{LO}}$ exhibit some intriguing properties, with the most significant one being their power behavior in relation to the variables $Q$ and $\delta$.

To illustrate the physical implications of these power behaviors clearly, we compare two cases in terms of their behaviors. In the first case, we examine the power behaviors of the contributions from one-$Z$ exchange and $\gamma Z$ exchange, considering only the LO interactions. In the second case, we compare the power behaviors of the imaginary parts of the contributions from $\gamma Z$ exchange, considering both the LO interactions and the NLO interactions.


When considering only the LO interactions, the couplings $F_2,g_2$ are zero, while $F_{1}$ and $g_{1,3}$ are nonzero. In this case, all interactions are well defined within the standard model. At the tree level, there are two nonzero PV contributions, namely, $\textrm{Re}[{\cal F}_{Z,1}^{V}]$ and $\textrm{Re}[{\cal F}_{Z,1}^{A}]$, which are independent of $Q$ and $\delta$.  Regarding the $\gamma Z$-exchange contributions, there are now six nonzero contributions, such as $C_{\gamma Z,i11}^{V}$ and $C_{\gamma Z,i13}^{A}$. By utilizing the analytic expressions given in Eqs. (\ref{eq-low-energy-expressions-Re-CV}), (\ref{eq-low-energy-expressions-Re-CA}), (\ref{eq-low-energy-expressions-Im-CV}), (\ref{eq-low-energy-expressions-Im-CA}), one can directly obtain the power behaviors of these contributions at low energy.

\begin{table}[htbp]
\centering
\begin{tabular}
{|p{2.5cm}<{\centering}|p{5cm}<{\centering}||p{2.5cm}<{\centering}|p{5cm}<{\centering}|}
\hline
& power behavior &  & power behavior  \\
\hline
$\textrm{Re}[{\cal F}_{Z,1}^{V}]$
& $1$
& $\textrm{Re}[C_{\gamma Z,111}^{V}]$
& $\alpha_e\frac{2M_NQ+\delta}{Q^2}$ \\
\hline
$\textrm{Re}[{\cal F}_{Z,2}^{V}]$
&no contribution
&$\textrm{Re}[C_{\gamma Z,211}^{V}]$
&$\alpha_e\frac{Q(2M_NQ+\delta)}{M_N^3}\log\frac{4M_N^4}{\nu^2-Q^4}$\\
\hline
$\textrm{Re}[{\cal F}_{Z,3}^{V}]$
&no contribution
&$\textrm{Re}[C_{\gamma Z,311}^{V}]$
&$\alpha_e\frac{M_N}{Q}$\\
\hline
$ \textrm{Re}[{\cal F}_{Z,1}^{A}]$
& $1$
& $\textrm{Re}[C_{\gamma Z,113}^{A}]$
& $\alpha_e\frac{2M_NQ+\delta}{Q^2}$ \\
\hline
$ \textrm{Re}[{\cal F}_{Z,2}^{A}]$
& no contribution
& $\textrm{Re}[C_{\gamma Z,213}^{A}]$
& $\alpha_e\frac{M_N}{Q}$ \\
\hline
$ \textrm{Re}[{\cal F}_{Z,3}^{A}]$
& no contribution
& $\textrm{Re}[C_{\gamma Z,313}^{A}]$
& $\alpha_e\frac{M_N}{Q}$ \\
\hline
\end{tabular}
\caption{The power behaviors of the contributions $\textrm{Re}[{\cal F}_{Z,i}^{V}]$, $\textrm{Re}[{\cal F}_{Z,i}^{A}]$, $\textrm{Re}[C_{\gamma Z,i11}^{V}]$, and $\textrm{Re}[C_{\gamma Z,i13}^{A}]$, where only the contributions from the LO interactions are considered and the contributions from $R_{\textrm{IR}}$ have been neglected.}
\label{Tab-low-energy-behavior-Re-FVA-vs-CVA-point-like}
\end{table}

For convenience, we present the power behaviors of $\textrm{Re}[{\cal F}{Z,i}^{V}]$, $\textrm{Re}[{\cal F}{Z,i}^{A}]$, $\textrm{Re}[C_{\gamma Z,i11}^{V}]$, and $\textrm{Re}[C_{\gamma Z,i13}^{A}]$ together in Table \ref{Tab-low-energy-behavior-Re-FVA-vs-CVA-point-like}. The comparisons clearly demonstrate that, except for $\textrm{Re}[C_{\gamma Z,211}^{V}]$, the coefficients from the $\gamma Z$-exchange contributions always exhibit an enhanced factor of $M_N/Q$ for any given $\delta$. This implies that when $Q/M_N \approx \alpha_e$, the $\gamma Z$ exchange can provide contributions of comparable magnitude to the one-$Z$ exchange.

This enhancement is not surprising. A similar property occurs in the pure electromagnetic system near the threshold, where contributions from multiple-photon exchange need to be summed. In the case of bound states governed by pure electromagnetic interactions, typically only the ladder diagrams are summed.

The detailed calculations demonstrate that the enhanced factor $M_N/Q$ appears not only in the sum of box diagrams [Fig. \ref{Figure-general-gammaZ-exchange-in-ep} (a+c)] but also in the sum of crossed box diagrams [Fig. \ref{Figure-general-gammaZ-exchange-in-ep} (b+d)]. This is in stark contrast to the case of pure electromagnetic interactions.

\begin{table}[htbp]
\centering
\begin{tabular}
{|p{2.5cm}<{\centering}|p{5cm}<{\centering}||p{2.5cm}<{\centering}|p{5cm}<{\centering}|}
\hline
& power behavior &  & power behavior  \\
\hline
$\textrm{Im}[{\cal F}_{Z,1}^{V}]$
& no contribution
& $\textrm{Im}[C_{\gamma Z,111}^{V}]$
& $\alpha_e\textrm{Max}\{\frac{z^2}{M_N^2Q^2},1\}$ \\
\hline
$\textrm{Im}[{\cal F}_{Z,2}^{V}]$
& no contribution
&$\textrm{Im}[C_{\gamma Z,211}^{V}]$
&$\alpha_e\frac{Q}{M_N}$\\
\hline
$\textrm{Im}[{\cal F}_{Z,3}^{V}]$
& no contribution
&$\textrm{Im}[C_{\gamma Z,311}^{V}]$
&$\alpha_e\frac{2M_NQ+\delta}{M_NQ}$\\
\hline
$ \textrm{Im}[{\cal F}_{Z,1}^{A}]$
& no contribution
& $\textrm{Im}[C_{\gamma Z,113}^{A}]$
& $\alpha_e\textrm{Max}\{\frac{z^2}{M_N^2Q^2},1\}$ \\
\hline
$ \textrm{Im}[{\cal F}_{Z,2}^{A}]$
& no contribution
& $\textrm{Im}[C_{\gamma Z,213}^{A}]$
& $\alpha_e\frac{2M_NQ+\delta}{M_NQ}$ \\
\hline
$ \textrm{Im}[{\cal F}_{Z,3}^{A}]$
& no contribution
& $\textrm{Im}[C_{\gamma Z,313}^{A}]$
& $\alpha_e\frac{2M_NQ+\delta}{M_NQ}$ \\
\hline
\end{tabular}
\caption{The power behaviors of the contributions $\textrm{Im}[{\cal F}_{Z,i}^{V}]$, $\textrm{Im}[{\cal F}_{Z,i}^{A}]$, $\textrm{Im}[C_{\gamma Z,i11}^{V}]$  and $\textrm{Im}[C_{\gamma Z,i13}^{A}]$, where only the contributions from the LO interactions are considered and the contributions from $I_{\textrm{IR}}$ have been neglected.}
\label{Tab-low-energy-behavior-Im-FVA-vs-CVA-point-like}
\end{table}
In Table \ref{Tab-low-energy-behavior-Im-FVA-vs-CVA-point-like}, we present the power behaviors of $\textrm{Im}[C_{\gamma Z,i11}^{V}]$ and $\textrm{Im}[C_{\gamma Z,i13}^{A}]$. The results clearly demonstrate that, in the region $\delta \gg M_NQ$, there is enhancement for all coefficients except $\textrm{Im}[C_{\gamma Z,211}^{V}]$, while no enhancement is observed in other regions. This behavior differs from that of the real parts. In literature, the imaginary parts are typically used as inputs in DRs to estimate the real parts of the $\gamma Z$-exchange contributions. Our results highlight the importance of carefully considering the imaginary parts of $C_{\gamma Z,ijk}^{V}$ and $C_{\gamma Z,ijk}^{A}$ even when considering only the leading-order (LO) interactions, particularly in the region where $\alpha_e\delta/M_NQ \sim 1$.

When considering interactions beyond the LO, the effective interactions with nonzero $F_2$ and $g_2$ come into play. In this case, at the low-energy scale, the corresponding radiative contributions should be combined with the four-fermion contact interactions to yield the final physical contributions. An important characteristic is that, at the one-loop level, the contact interactions do not alter the imaginary parts of the $\gamma Z$-exchange contributions. This implies that the low-energy behaviors of the imaginary parts are physical, and they can serve as a unique means to verify the power counting rules.

In Tables \ref{Tab-low-energy-behavior-Im-CV-ratio-LO-and-NLO} and \ref{Tab-low-energy-behavior-Im-CA-ratio-LO-and-NLO}, we present the power behaviors of the ratios $\textrm{Im}[C_{\gamma Z,ijk}^{V}]/\textrm{Im}[C_{\gamma Z,i11}^{V}]$ and $\textrm{Im}[C_{\gamma Z,i23}^{A}]/\textrm{Im}[C_{\gamma Z,i13}^{A}]$ in different regions. The results indicate that the naive NNLO contributions, $\textrm{Im}[C_{\gamma Z,i22}^{V}]$, are of higher order, while the naive NLO contributions are actually of the same order as the LO contributions. These observations suggest that the naive power counting rules are not preserved in certain cases.

\begin{table}[htbp]
\renewcommand{\arraystretch}{1.0}
\centering
\begin{tabular}
{|p{5cm}<{\centering} p{2.5cm}<{\centering} p{2.5cm}<{\centering} p{2.5cm}<{\centering}|}
\hline
power behavior & $\delta\ll M_NQ$ & $\delta \approx M_NQ$  & $\delta \gg M_NQ$  \\
\hline
$\textrm{Im}[C_{\gamma Z,112}^{V}]/\textrm{Im}[C_{\gamma Z,111}^{V}]$
& $1$
& $1$
& $1$ \\

$\textrm{Im}[C_{\gamma Z,121}^{V}]/\textrm{Im}[C_{\gamma Z,111}^{V}]$
& $1$
& $1$
& $1$ \\

$\textrm{Im}[C_{\gamma Z,122}^{V}]/\textrm{Im}[C_{\gamma Z,111}^{V}]$
& $\frac{Q}{M_N}$
& $\frac{Q}{M_N}$
& $\frac{\delta^2}{M_N^2}$
\\
\hline
\hline
$\textrm{Im}[C_{\gamma Z,212}^{V}]/\textrm{Im}[C_{\gamma Z,211}^{V}]$
& $1$
& $1$
& $1$
\\

$\textrm{Im}[C_{\gamma Z,212}^{V}]/\textrm{Im}[C_{\gamma Z,211}^{V}]$
& $1$
& $1$
& $1$
\\

$\textrm{Im}[C_{\gamma Z,222}^{V}]/\textrm{Im}[C_{\gamma Z,211}^{V}]$
& $O(M_Z^{-2})$
& $O(M_Z^{-2})$
& $O(M_Z^{-2})$
 \\
\hline
\hline
$\textrm{Im}[C_{\gamma Z,312}^{V}]/\textrm{Im}[C_{\gamma Z,311}^{V}]$
& $1$
& $1$
& $1$
 \\
$\textrm{Im}[C_{\gamma Z,321}^{V}]/\textrm{Im}[C_{\gamma Z,311}^{V}]$
& $1$
& $1$
& $1$
 \\
$\textrm{Im}[C_{\gamma Z,322}^{V}]/\textrm{Im}[C_{\gamma Z,311}^{V}]$
& $\frac{Q}{M_N}$
& $\frac{Q}{M_N}$
& $\frac{\delta}{M_N^2}$ \\
\hline
\end{tabular}
\caption{The power behaviors of the ratios $\textrm{Im}[C_{\gamma Z,ijk}^{V}]/\textrm{Im}[C_{\gamma Z,i11}^{V}]$.}
\label{Tab-low-energy-behavior-Im-CV-ratio-LO-and-NLO}
\end{table}

\begin{table}[htb]
\centering
\begin{tabular}
{|p{5cm}<{\centering} p{2.5cm}<{\centering} p{2.5cm}<{\centering} p{2.5cm}<{\centering}|}
\hline
power behavior & $\delta\ll M_NQ$ & $\delta \approx M_NQ$  & $\delta \gg M_NQ$  \\
\hline
$\textrm{Im}[C_{\gamma Z,123}^{A}]/\textrm{Im}[C_{\gamma Z,113}^{A}]$
& $1$
& $1$
& $1$ \\
\hline
$\textrm{Im}[C_{\gamma Z,223}^{A}]/\textrm{Im}[C_{\gamma Z,213}^{A}]$
& $1$
& $1$
& $1$ \\
\hline
$\textrm{Im}[C_{\gamma Z,323}^{A}]/\textrm{Im}[C_{\gamma Z,313}^{A}]$
& $1$
& $1$
& $1$ \\
\hline
\end{tabular}
\caption{The power behaviors of the contributions $\textrm{Im}[C_{\gamma Z,i23}^{A}]/\textrm{Im}[C_{\gamma Z,i13}^{A}]$.}
\label{Tab-low-energy-behavior-Im-CA-ratio-LO-and-NLO}
\end{table}

\subsection{Power behaviors of the helicity amplitudes}
Since the helicity amplitudes directly correspond to observable physical quantities, their properties may provide a more definite reflection of the physical meaning compared to the coefficients $C_{\gamma Z,ijk}^{X}$. In this section, we present the power behaviors of the helicity amplitudes based on the expressions listed in Appendix B.

\begin{table}[htbp]
\renewcommand{\arraystretch}{1.0}
\centering
\begin{tabular}
{|p{2.5cm}<{\centering} p{5cm}<{\centering}|p{2.5cm}<{\centering} p{5cm}<{\centering}|}
\hline
& power behavior &  & power behavior  \\
\hline
$\textrm{Re}[{\cal M}_{Z,1}^{++++,V}]$
& $hz^2 $
& $\textrm{Re}[{\cal M}_{\gamma Z,11}^{++++,V}]$
& $\alpha_{e}h \textrm{Max}\{M_N^2Q^2,z^2\}$  \\
$\textrm{Re}[{\cal M}_{Z,3}^{++++,A}]$
& $h(8M_N^2Q^2+z^2)$
& $\textrm{Re}[{\cal M}_{\gamma Z,13}^{++++,A}]$
& $\alpha_{e}hM_N^2(2M_NQ+\delta)$ \\
\hline
\hline
$\textrm{Re}[{\cal M}_{Z,1}^{+++-,V}]$
&$hzM_NQ$
&$\textrm{Re}[{\cal M}_{\gamma Z,11}^{+++-,V}]$
&$\alpha_{e}hzM_NQ$\\
$\textrm{Re}[{\cal M}_{Z,3}^{+++-,A}]$
& $hzM_NQ$
& $\textrm{Re}[{\cal M}_{\gamma Z,13}^{+++-,A}]$
& $\alpha_{e}hz\frac{M_N(2M_NQ+\delta)}{Q}$ \\
\hline
\hline
$\textrm{Re}[{\cal M}_{Z,1}^{++--,V}]$
&$hz^2$
&$\textrm{Re}[{\cal M}_{\gamma Z,11}^{++--,V}]$
&$\alpha_{e}hz^2$\\
$\textrm{Re}[{\cal M}_{Z,3}^{++--,A}]$
& $hz^2$
& $\textrm{Re}[{\cal M}_{\gamma Z,13}^{++--,A}]$
& $\alpha_{e}hz^2$ \\
\hline
\end{tabular}
\caption{The power behaviors of $\textrm{Re}[{\cal M}_{Z,1}^{\pm\pm\pm\pm,V}]$, $\textrm{Re}[{\cal M}_{Z,3}^{\pm\pm\pm\pm,A}]$, $\textrm{Re}[{\cal M}_{\gamma Z,11}^{\pm\pm\pm\pm,V}]$, and $\textrm{Re}[{\cal M}_{\gamma Z,13}^{\pm\pm\pm\pm,V}]$, where only the LO interactions are considered and the contributions from $R_{\textrm{IR}}$ have been neglected.}
\label{Tab-low-energy-behavior-Re-Helicity-Amplitudes-LO-interactions}
\end{table}

\begin{table}[htbp]
\renewcommand{\arraystretch}{1.0}
\centering
\begin{tabular}
{|p{2.5cm}<{\centering} p{5cm}<{\centering}|p{2.5cm}<{\centering} p{5cm}<{\centering}|}
\hline
& power behavior &  & power behavior
\\
\hline
$\textrm{Im}[{\cal M}_{Z,1}^{++++,V}]$
& no contribution
& $\textrm{Im}[{\cal M}_{\gamma Z,11}^{++++,V}]$
& $\alpha_{e}hz^2$
\\
$\textrm{Im}[{\cal M}_{Z,3}^{++++,A}]$
& no contribution
& $\textrm{Im}[{\cal M}_{\gamma Z,13}^{++++,A}]$
& $\alpha_{e}h(8M_N^2Q^2+3z^2)$
\\
\hline
\hline
$\textrm{Im}[{\cal M}_{Z,1}^{+++-,V}]$
& no contribution
&$\textrm{Im}[{\cal M}_{\gamma Z,11}^{+++-,V}]$
&$\alpha_{e}hzM_NQ$
\\
$\textrm{Im}[{\cal M}_{Z,3}^{+++-,A}]$
& no contribution
& $\textrm{Im}[{\cal M}_{\gamma Z,13}^{+++-,A}]$
& $\alpha_{e}hz\frac{(3M_NQ+\delta)(M_NQ+\delta)}{M_NQ}$
\\
\hline
\hline
$\textrm{Im}[{\cal M}_{Z,1}^{++--,V}]$
& no contribution
&$\textrm{Im}[{\cal M}_{\gamma Z,11}^{++--,V}]$
&$\alpha_{e}hz^2$
\\
$\textrm{Im}[{\cal M}_{Z,3}^{++--,A}]$
& no contribution
& $\textrm{Im}[{\cal M}_{\gamma Z,13}^{++--,A}]$
& $\alpha_{e}hz^2$ \\
\hline
\end{tabular}
\caption{The power behaviors of $\textrm{Im}[{\cal M}_{Z,1}^{\pm\pm\pm\pm,V}]$, $\textrm{Im}[{\cal M}_{Z,3}^{\pm\pm\pm\pm,A}]$, $\textrm{Im}[{\cal M}_{\gamma Z,11}^{\pm\pm\pm\pm,V}]$, and $\textrm{Im}[{\cal M}_{\gamma Z,13}^{\pm\pm\pm\pm,V}]$, where only the LO interactions are considered.}
\label{Tab-low-energy-behavior-Im-Helicity-Amplitudes-LO-interactions}
\end{table}

\begin{table}[htbp]
\centering
\begin{tabular}
{|p{7cm}<{\centering} p{2cm}<{\centering} p{2cm}<{\centering} p{2cm}<{\centering}|}
\hline
power behaviors in different regions & $\delta \ll M_NQ$ & $\delta\approx  M_NQ$  & $\delta \gg M_NQ$  \\
\hline
$\textrm{Re}[{\cal M}_{\gamma Z,11}^{++++,V}]/\textrm{Re}[{\cal M}_{Z,1}^{++++,V}]$
& $\alpha_{e} \uwave{\frac{M_NQ}{\delta}}$
& $\alpha_{e}$
& $\alpha_{e}$  \\
$\textrm{Re}[{\cal M}_{\gamma Z,13}^{++++,A}]/\textrm{Re}[{\cal M}_{Z,3}^{++++,A}]$
& $\alpha_{e}\uwave{\frac{M_N}{Q}}$
& $\alpha_{e}\uwave{\frac{M_N}{Q}}$
& $\alpha_{e}\uwave{\frac{M_N^2}{\delta}}$ \\
\hline
\hline
$\textrm{Re}[{\cal M}_{\gamma Z,11}^{+++-,V}]/\textrm{Re}[{\cal M}_{Z,1}^{+++-,V}]$
& $\alpha_{e}$
& $\alpha_{e}$
& $\alpha_{e}$\\
$\textrm{Re}[{\cal M}_{\gamma Z,13}^{A,+++-}]/\textrm{Re}[{\cal M}_{Z,3}^{A,+++-}]$
& $\alpha_{e}\uwave{\frac{M_N}{Q}}$
& $\alpha_{e}\uwave{\frac{M_N}{Q}}$
& $\alpha_{e}\uwave{\frac{\delta}{Q^2}}$ \\
\hline
\hline
$\textrm{Re}[{\cal M}_{\gamma Z,11}^{++--,V}]/\textrm{Re}[{\cal M}_{Z,1}^{++--,V}]$
&$\alpha_{e}$
&$\alpha_{e}$
&$\alpha_{e}$\\
$\textrm{Re}[{\cal M}_{\gamma Z,13}^{A,++--}]/\textrm{Re}[{\cal M}_{Z,3}^{A,++--}]$
& $\alpha_{e}$
& $\alpha_{e}$
& $\alpha_{e}$ \\
\hline
\end{tabular}
\caption{The power behaviors of $\textrm{Re}[{\cal M}_{\gamma Z,1k}^{\pm\pm\pm\pm,X}]/\textrm{Re}[{\cal M}_{Z,k}^{\pm\pm\pm\pm,X}]$ in different regions, where only the LO interactions are considered. The terms with enhanced factors are labeled by a wavy line.}
\label{Tab-low-energy-behavior-Re-Helicity-Amplitudes-ratio-LO-interactions}
\end{table}

In Tables \ref{Tab-low-energy-behavior-Re-Helicity-Amplitudes-LO-interactions} and \ref{Tab-low-energy-behavior-Im-Helicity-Amplitudes-LO-interactions}, we present the power behaviors of the helicity amplitudes resulting from one-$Z$ exchange and $\gamma Z$ exchange, considering only the LO interactions. In Table  \ref{Tab-low-energy-behavior-Re-Helicity-Amplitudes-ratio-LO-interactions}, we list the power behaviors of the ratios $\textrm{Re}[{\cal M}{\gamma Z,1k}^{\pm\pm\pm\pm,X}]/\textrm{Re}[{\cal M}{Z,k}^{\pm\pm\pm\pm,X}]$ in different regions, where the enhanced factors are indicated by a wavy line. These enhancements suggest that, even when considering only the LO interactions, diagrams with higher orders of $\alpha_e$ should be considered and summed to obtain the correct contributions in specific regions. Directly estimating the $\gamma Z$-exchange contributions to the helicity amplitudes through loop integrals or dispersion relations is only valid outside these regions. The results in Table \ref{Tab-low-energy-behavior-Re-Helicity-Amplitudes-ratio-LO-interactions} clearly reveal the availability of specific $\gamma Z$-exchange helicity amplitudes in different regions. Combining all these regions, we find that only in the region, where $\alpha_e\ll Q/M_N\sim \delta/M_N^2\ll 1$, all the $\gamma Z$-exchange helicity amplitudes are applicable. Outside this region, higher-order radiative corrections should be taken into account.

The full physical helicity amplitudes of $ep$ scattering are the linear sum of the $V$ parts and the $A$ parts. Therefore, when considering only the LO interactions, the corresponding ratios of the full physical helicity amplitudes can be expressed as
\begin{eqnarray}
\frac{\textrm{Re}[{\cal M}_{\gamma Z,11}^{\pm\pm\pm\pm,V}]F_1g_1g_e^A+\textrm{Re}[{\cal M}_{\gamma Z,13}^{\pm\pm\pm\pm,A}]F_1g_3g_e^V}{\textrm{Re}[{\cal M}_{ Z,1}^{\pm\pm\pm\pm,V}]g_1g_e^A+\textrm{Re}[{\cal M}_{ Z,3}^{\pm\pm\pm\pm,A}]g_3g_e^V}. \nonumber
\end{eqnarray}
By combining the power behaviors listed in Tables \ref{Tab-low-energy-behavior-Re-Helicity-Amplitudes-LO-interactions} and \ref{Tab-low-energy-behavior-Re-Helicity-Amplitudes-ratio-LO-interactions}, one can observe that there are still enhancements for the full physical helicity amplitudes in the $++++$ and $+++-$ cases when assuming $g_1g_{e}^{A}$ and $g_3g_e^{V}$ to be at the same order.

\begin{table}[htbp]
\centering
\begin{tabular}
{|p{7cm}<{\centering} p{2cm}<{\centering} p{2cm}<{\centering} p{2cm}<{\centering}|}
\hline
power behaviors in different regions & $\delta \ll M_NQ$ & $\delta\approx  M_NQ$  & $\delta \gg M_NQ$  \\
\hline
$\textrm{Im}[{\cal M}_{\gamma Z,12}^{++++,V}]/\textrm{Im}[{\cal M}_{\gamma Z,11}^{++++,V}]$
& $\uwave{\frac{Q^2}{\delta}}$
& $\frac{Q}{M_N}$
& $\frac{\delta}{M_N^2}$
\\
$\textrm{Im}[{\cal M}_{\gamma Z,21}^{++++,V}]/\textrm{Im}[{\cal M}_{\gamma Z,11}^{++++,V}]$
& $\uwave{\frac{Q^2}{\delta}}$
& $\frac{Q}{M_N}$
& $\frac{\delta}{M_N^2}$
\\
$\textrm{Im}[{\cal M}_{\gamma Z,22}^{++++,V}]/\textrm{Im}[{\cal M}_{\gamma Z,11}^{++++,V}]$
& $\frac{Q^3}{M_N\delta}$
& $\frac{Q^2}{M_N^2}$
& $\frac{\delta^2}{M_N^4}$
\\

$\textrm{Im}[{\cal M}_{\gamma Z,23}^{++++,A}]/\textrm{Im}[{\cal M}_{\gamma Z,13}^{++++,A}]$
& $\uwave{1}$
& $\uwave{1}$
& $\uwave{1}$ \\
\hline
\hline
$\textrm{Im}[{\cal M}_{\gamma Z,12}^{+++-,V}]/\textrm{Im}[{\cal M}_{Z,11}^{+++-,V}]$
& $\frac{Q}{M_N}$
& $\frac{Q}{M_N}$
& $\frac{\delta}{M_N^2}$
\\
$\textrm{Im}[{\cal M}_{\gamma Z,21}^{+++-,V}]/\textrm{Im}[{\cal M}_{Z,11}^{+++-,V}]$
& $\frac{Q}{M_N}$
& $\frac{Q}{M_N}$
& $\frac{\delta}{M_N^2}$
\\

$\textrm{Im}[{\cal M}_{\gamma Z,22}^{+++-,V}]/\textrm{Im}[{\cal M}_{Z,11}^{+++-,V}]$
& $\frac{Q^2}{M_N^2}$
& $\frac{Q^2}{M_N^2}$
& $\frac{\delta^2}{M_N^4}$\\

$\textrm{Im}[{\cal M}_{\gamma Z,23}^{A,+++-}]/\textrm{Im}[{\cal M}_{Z,13}^{A,+++-}]$
& $\uwave{1}$
& $\uwave{1}$
& $\uwave{1}$ \\
\hline
\hline
$\textrm{Im}[{\cal M}_{\gamma Z,12}^{++--,V}]/\textrm{Im}[{\cal M}_{Z,11}^{++--,V}]$
&$\frac{Q}{M_N}$
&$\frac{Q}{M_N}$
&$\frac{\delta}{M_N^2}$
\\
$\textrm{Im}[{\cal M}_{\gamma Z,12}^{++--,V}]/\textrm{Im}[{\cal M}_{Z,11}^{++--,V}]$
&$\frac{Q}{M_N}$
&$\frac{Q}{M_N}$
&$\frac{\delta}{M_N^2}$
\\

$\textrm{Im}[{\cal M}_{\gamma Z,22}^{++--,V}]/\textrm{Im}[{\cal M}_{Z,11}^{++--,V}]$
& $\frac{Q^2}{M_N^2}$
& $\frac{Q^2}{M_N^2}$
& $\frac{Q^2}{M_N^2}$
\\

$\textrm{Im}[{\cal M}_{\gamma Z,23}^{A,++--}]/\textrm{Im}[{\cal M}_{Z,13}^{A,++--}]$
&$\frac{Q}{M_N}$
&$\frac{Q}{M_N}$
&$\frac{\delta}{M_N^2}$\\
\hline
\end{tabular}
\caption{The power behaviors of $\textrm{Im}[{\cal M}_{\gamma Z,jk}^{\pm\pm\pm\pm,X}]/\textrm{Re}[{\cal M}_{\gamma Z,1n}^{\pm\pm\pm\pm,X}]$ in different regions, where the contributions at the same order are labeled by a wavy line.}
\label{Tab-low-energy-behavior-Im-Helicity-Amplitudes-ratio-NLO-interactions}
\end{table}

In Table \ref{Tab-low-energy-behavior-Im-Helicity-Amplitudes-ratio-NLO-interactions}, we present the power behaviors of the ratios $\textrm{Im}[{\cal M}_{\gamma Z,ij}^{\pm\pm\pm\pm,V}]/\textrm{Im}[{\cal M}_{\gamma Z,11}^{\pm\pm\pm\pm,V}]$ and $\textrm{Im}[{\cal M}_{\gamma Z,ij}^{\pm\pm\pm\pm,A}]/\textrm{Im}[{\cal M}_{\gamma Z,13}^{\pm\pm\pm\pm,A}]$ in different regions. These results show that, in the region $\delta\leq Q^2 \ll M_NQ$, the contributions from the NLO interactions, $\textrm{Im}[{\cal M}_{\gamma Z,12}^{++++,V}]$ and $\textrm{Im}[{\cal M}_{\gamma Z,21}^{++++,V}]$, are even larger than the contribution from the LO interactions, $\textrm{Im}[{\cal M}_{\gamma Z,11}^{++++,V}]$. Furthermore, for any $\delta$, the contributions from the NLO interactions, $\textrm{Im}[{\cal M}_{\gamma Z,23}^{++++,A}]$ or $\textrm{Im}[{\cal M}_{\gamma Z,23}^{+++-,A}]$, are at the same order as the contributions from the LO interactions, $\textrm{Im}[{\cal M}_{\gamma Z,13}^{++++,A}]$ or $\textrm{Im}[{\cal M}_{\gamma Z,23}^{++++,A}]$, respectively. Since the imaginary parts of these contributions cannot be canceled by contact interactions, these properties indicate that the naive power counting rules are broken in these cases. On the other hand, contributions involving higher interactions, such as $\textrm{Im}[{\cal M}_{\gamma Z,22}^{\pm\pm\pm\pm,V}]$, are much smaller and follow the naive power counting rules.

In practical calculations, we also considered interactions involving higher-order momentum, such as $F_1q_\gamma^2$ (where $q_\gamma$ is the momentum of the incoming photon), to examine their behavior. We found that the imaginary parts of the corresponding contributions are at higher orders. This indicates that, although certain NLO contributions break the naive power counting rules, the imaginary parts of contributions ${\cal M}_{ij}^{(a+b+c+d)}$ with $i+j>3$ (such as the NNLO interactions) can be safely neglected in the low energy limit. These important observations suggest that, although the naive power counting rules for the imaginary parts are not upheld, there are still regular power rules governing the imaginary parts of the contributions.

Due to these power behaviors, we conclude that in the low-energy regions where the radiative corrections are not strongly enhanced, the imaginary parts are reliable when both the LO and the NLO interactions are included. This also means that the corresponding real parts are reliable since the real and imaginary parts obtained in our calculation obey the DRs. The power behaviors of the imaginary parts also suggest a systematic way to estimate the $\gamma Z$-exchange contributions to higher orders of low energy: one can take the effective interactions with higher order momentum as inputs to obtain the corresponding imaginary parts of the amplitudes and then use the DRs to obtain the corresponding real parts. Such a method can avoid the breakdown of the power counting rules that occur in the real parts. Naturally, the cost is that some unknown constants may be introduced to absorb the contributions from high energy.

\subsection{Power behaviors of ${\cal A}_{\gamma Z,ijk}^{V}$ and ${\cal A}_{\gamma Z,ijk}^{A}$}

In Table \ref{Tab-low-energy-behavior-Re-APV-LO-interactions}, we present the power behaviors of $\textrm{Re}[{\cal A}{Z,11}^{V}]$, $\textrm{Re}[{\cal A}{Z,13}^{A}]$, $\textrm{Re}[{\cal A}{\gamma Z,111}^{V}]$, and $\textrm{Re}[{\cal A}{\gamma Z,113}^{A}]$, where only the LO interactions are considered. The results clearly show that the $\gamma Z$-exchange contribution $\textrm{Re}[{\cal A}{\gamma Z,111}^{V}]$ is always smaller than the one-$Z$-exchange contribution $\textrm{Re}[{\cal A}{Z,11}^{V}]$. In the region with $\delta\approx M_NQ$ or $\delta\gg M_NQ$, the $\gamma Z$-exchange contribution $\textrm{Re}[{\cal A}{\gamma Z,113}^{A}]$ is as large as the one-$Z$-exchange contribution $\textrm{Re}[{\cal A}{Z,13}^{A}]$, but it is still much smaller than the one-$Z$-exchange contribution $\textrm{Re}[{\cal A}{Z,11}^{V}]$ in these two regions. Combining these properties, one can conclude that there is no additional enhancement in the $\gamma Z$-exchange contributions to the full $A{\textrm{PV}}$. This is very different from the properties of $\gamma Z$-exchange contributions to the helicity amplitudes or the coefficients.

\begin{table}[htbp]
\centering
\begin{tabular}
{|p{4cm}<{\centering} p{2.5cm}<{\centering} p{4cm}<{\centering} p{2cm}<{\centering} p{2cm}<{\centering}|}
\hline
power behavior & $\delta\ll Q^3/M_N$ & $ Q^3/M_N\ll\delta \ll M_NQ$ & $\delta \approx  M_NQ$  & $\delta \gg M_NQ$ \\
\hline
${\textrm{Re}[\cal A}_{Z,11}^{V}]$
& $M_N^2Q^6 $
& $M_N^3Q^3\delta $
& $M_N^4Q^4$
& $M_N^2Q^2\delta^2$  \\
$\textrm{Re}[{\cal A}_{\gamma Z,111}^{V}]$
& $\alpha_eM_N^3Q^3\delta$
& $\alpha_eM_N^3Q^3\delta$
& $\alpha_eM_N^4Q^4$
& $\alpha_eM_N^2Q^2\delta^2$\\
\hline
\hline
$\textrm{Re}[{\cal A}_{Z,13}^{A}]$
& $M_N^3Q^5$
& $M_N^3Q^5$
& $M_N^3Q^5$
& $M_N^2Q^4\delta$ \\
$\textrm{Re}[{\cal A}_{\gamma Z,113}^{A}]$
& $\alpha_eM_N^3Q^5$
& $\alpha_eM_N^3Q^5$
& $\alpha_e\uwave{M_N^4Q^4}$
& $\alpha_e\uwave{M_N^2Q^2\delta^2}$\\
\hline
\end{tabular}
\caption{The power behaviors of $\textrm{Re}[{\cal A}_{Z,11}^{V}]$, $\textrm{Re}[{\cal A}_{Z,13}^{A}]$, $\textrm{Re}[{\cal A}_{\gamma Z,111}^{V}]$, and $\textrm{Re}[{\cal A}_{\gamma Z,113}^{A}]$, where only the LO interactions are considered. The relative enhanced terms are labeled by a wavy line.}
\label{Tab-low-energy-behavior-Re-APV-LO-interactions}
\end{table}

\begin{table}[htbp]
\centering
\begin{tabular}
{|p{4cm}<{\centering} p{2cm}<{\centering} p{2cm}<{\centering} p{2cm}<{\centering}|}
\hline
power behaviors& $\delta \ll M_NQ$ & $\delta\approx  M_NQ$  & $\delta \gg M_NQ$  \\
\hline
$\textrm{Im}[{\cal A}_{\gamma Z,112}^{V}]/\textrm{Im}[{\cal A}_{\gamma Z,111}^{V}]$
& $\frac{Q^3}{M_N\delta}$
& $\frac{Q^2}{M_N^2}$
& $\frac{\delta^2}{M_N^4}$  \\

$\textrm{Im}[{\cal A}_{\gamma Z,121}^{V}]/\textrm{Im}[{\cal A}_{\gamma Z,111}^{V}]$
& $\frac{Q^3}{M_N\delta}$
& $\frac{Q^2}{M_N^2}$
& $\frac{\delta^2}{M_N^4}$  \\

$\textrm{Im}[{\cal A}_{\gamma Z,122}^{V}]/\textrm{Im}[{\cal A}_{\gamma Z,111}^{V}]$
& $\frac{Q^2}{M_N^2}$
& $\frac{Q^2}{M_N^2}$
& $\frac{\delta^2}{M_N^4}$  \\

$\textrm{Im}[{\cal A}_{\gamma Z,211}^{V}]/\textrm{Im}[{\cal A}_{\gamma Z,111}^{V}]$
& $\frac{Q^3}{M_N\delta}$
& $\frac{Q^2}{M_N^2}$
& $\frac{Q^2}{M_N^2}$  \\

$\textrm{Im}[{\cal A}_{\gamma Z,212}^{V}]/\textrm{Im}[{\cal A}_{\gamma Z,111}^{V}]$
& $\frac{Q^3}{M_N\delta}$
& $\frac{Q^2}{M_N^2}$
& $\frac{Q^2}{M_N^2}$  \\

$\textrm{Im}[{\cal A}_{\gamma Z,221}^{V}]/\textrm{Im}[{\cal A}_{\gamma Z,111}^{V}]$
& $\frac{Q^3}{M_N\delta}$
& $\frac{Q^2}{M_N^2}$
& $\frac{Q^2}{M_N^2}$  \\

$\textrm{Im}[{\cal A}_{\gamma Z,222}^{V}]/\textrm{Im}[{\cal A}_{\gamma Z,111}^{V}]$
& $\frac{Q^4}{M_N^2\delta}$
& $\frac{Q^3}{M_N^3}$
& $\frac{Q^2\delta}{M_N^4}$  \\

\hline
\hline

$\textrm{Im}[{\cal A}_{\gamma Z,123}^{A}]/\textrm{Im}[{\cal A}_{\gamma Z,113}^{A}]$
& $\frac{\delta}{M_NQ}$
& $1$
& $1$ \\

$\textrm{Im}[{\cal A}_{\gamma Z,213}^{A}]/\textrm{Im}[{\cal A}_{\gamma Z,113}^{A}]$
& $1$
& $1$
& $\frac{M_N^2Q^2}{\delta^2}$\\

$\textrm{Im}[{\cal A}_{\gamma Z,223}^{A}]/\textrm{Im}[{\cal A}_{\gamma Z,113}^{A}]$
& $\frac{Q}{M_N}$
& $\frac{Q}{M_N}$
& $\frac{Q^2}{\delta}$\\

\hline
\end{tabular}
\caption{The power behaviors of $\textrm{Im}[{\cal A}_{\gamma Z,ijk}^{V}]/\textrm{Im}[{\cal A}_{\gamma Z,111}^{V}]$ and $\textrm{Im}[{\cal A}_{\gamma Z,ijk}^{A}]/\textrm{Im}[{\cal A}_{\gamma Z,113}^{A}]$ in different regions.}
\label{Tab-low-energy-behavior-Im-APV-NLO-interactions}
\end{table}

In Table \ref{Tab-low-energy-behavior-Im-APV-NLO-interactions}, we present the power behaviors of the ratios $\textrm{Im}[{\cal A}{\gamma Z,ijk}^{V}]/\textrm{Im}[{\cal A}{\gamma Z,111}^{V}]$ and $\textrm{Im}[{\cal A}{\gamma Z,ijk}^{A}]/\textrm{Im}[{\cal A}{\gamma Z,113}^{A}]$. The results show that only the contributions $\textrm{Im}[{\cal A}{\gamma Z,123}^{A}]$ and $\textrm{Im}[{\cal A}{\gamma Z,213}^{A}]$ are at the same order as the contribution $\textrm{Im}[{\cal A}{\gamma Z,113}^{A}]$, and other contributions from the NLO interactions are always smaller than those from the LO interactions. This means that the naive power counting rules for $A{\textrm{PV}}$ are broken in some cases, and the contributions from the NLO interactions should be considered.

\section{Application}

As an application, one can directly apply the above results to discuss the $\gamma Z$-exchange contributions for the upcoming P2 experiment with $Q^2=0.0045$ GeV$^2$ and $E_e=0.155$ GeV. Since the terms ${\cal A}_{\gamma Z,122}^{V}$ and ${\cal A}_{\gamma Z,222}^{V}$ contain UV divergences and should be absorbed by certain contact terms, we neglect them in the current analysis. By taking the IR scale $\mu_{\textrm{IR}}=1$ GeV, the finite parts of the analytical expressions for $A_{\textrm{PV}}$ give the following results
\begin{eqnarray}
\Box_{\gamma Z}^{V,\textrm{LO}}(\textrm{P2})&=&10^{-4}\times\frac{g_e^{A}}{\sigma}(257.245F_1^2g_1+8.350F_1^2g_2+14.711 F_1F_2g_1+4.099F_2^2g_1+4.099F_2F_1g_2), \nonumber\\
\Box_{\gamma Z}^{A,\textrm{LO}}(\textrm{P2})&=&10^{-4}\times\frac{g_e^{V}}{\sigma}(414.330F_1^2+418.543F_1F_2-0.156F_2^2)g_3,
\end{eqnarray}
where we have used $g_i$ and $g_{e}^{V,A}$ while not $\bar{g}_i$ and $\bar{g}_{e}^{V,A}$ to express the results, and the terms with $\frac{1}{\widetilde{\epsilon}_{\textrm{IR}}}$ have been neglected.

For comparison, the direct numeric calculation of the full expressions gives the following results:
\begin{eqnarray}
\Box_{\gamma Z}^{V}(\textrm{P2})&=&10^{-4}\times\frac{g_e^{A}}{\sigma}(230.269F_1^2g_1+7.582F_1^2g_2+13.928 F_1F_2g_1+4.090F_2^2g_1+4.394F_2F_1g_2), \nonumber\\
\Box_{\gamma Z}^{A}(\textrm{P2})&=&10^{-4}\times\frac{g_e^{V}}{\sigma}(410.700F_1^2+410.961F_1F_2-0.115F_2^2)g_3.
\end{eqnarray}
We can observe that there is about $10\%$ difference between $\Box_{\gamma Z}^{V,\textrm{LO}}$ and $\Box_{\gamma Z}^{V}$, this is natural and acceptable, since for P2 experiment, the corresponding $\delta$ is about $0.45$GeV$^2$ which is not a small value.

\section{Conclusion}
In summary, the full results reveal many interesting and important properties of the $\gamma Z$-exchange contributions at the amplitude level, which are not evident in the contributions to the physical quantity $A_{\textrm{PV}}$. Our findings suggest that when $Q$ and $\delta$ approach 0 independently, it is important to carefully consider the $\gamma Z$-exchange contributions. Specifically, to estimate the $\gamma Z$-exchange contributions, both the LO and NLO interactions should be included, which goes beyond the naive power counting rules. Additionally, in the region with $\alpha_{e}\ll Q/M_N \sim \delta/M_N^2 \ll1$, our results can be applied to estimate the $\gamma Z$-exchange contributions to any related physical quantities. However, outside this region, for some helicity amplitudes, the higher-order radiative contributions should be considered and summed. For the full physical quantity $A_{\textrm{PV}}$, the expressions can be applied in a wider region with $(Q/M_N \ll1) \cap (\delta/M_N^2\ll1)$. Finally, for the imaginary parts, although some contributions from the NLO interactions are not suppressed by the factor $Q/M_N$, the contributions with more higher-order interactions are suppressed. As a result, the corresponding real parts can be estimated order by order at the low energy scale via the DRs, and the imaginary parts by the effective interactions.

\section{Acknowledgments}
H.Q.Z. would like to thank Shin-Nan Yang and Zhi-Hui Guo for their helpful suggestions and discussions. This work is funded in part by the National Natural Science Foundations of China under Grant Nos. 12075058, 12047503, and 11975075.

\section{Appendix}

\subsection{Expressions for $C_{\gamma Z,ijk}^{X,\textrm{LO}}$}
\setcounter{equation}{0}
\renewcommand\theequation{A.\arabic{equation}}

The nonzero LO contributions, $\textrm{Re}[C_{\gamma Z,ijk}^{V,\textrm{LO}}]$ and $\textrm{Re}[C_{\gamma Z,ijk}^{A,\textrm{LO}}]$, can be expressed as
\begin{eqnarray}
\textrm{Re}[C_{\gamma Z,111}^{V,\textrm{LO}}]&=& -\frac{\alpha_{e}}{\pi M_Z^2 }\frac{2M_NQ+\delta}{Q^2}\Big[\frac{5}{4}+3\log\frac{M_Z}{M_N}+\frac{Q^2}{2M_NQ+\delta}R_{\textrm{IR}}\Big], \nonumber\\
\textrm{Re}[C_{\gamma Z,112}^{V,\textrm{LO}}]&=&-\frac{\alpha_{e}}{\pi M_Z^2}\frac{2M_NQ+\delta}{Q^2}\Big[\frac{1}{2}+2\log\frac{M_Z}{M_N}\Big],  \nonumber\\
\textrm{Re}[C_{\gamma Z,121}^{V,\textrm{LO}}]&=&\textrm{Re}[C_{\gamma Z,112}^{V,\textrm{LO}}], \nonumber\\
\textrm{Re}[C_{\gamma Z,122}^{V,\textrm{LO}}]&=&\frac{\alpha_{e}}{\pi M_Z^2 }\frac{2M_NQ+\delta}{Q^2}\Big[\frac{1}{4}+\frac{9M_Z^2}{16M_N^2}-\log\frac{M_Z}{M_N}+\frac{3M_Z^2}{8M_N^2}R_{\textrm{UV}}\Big], \nonumber\\
\textrm{Re}[C_{\gamma Z,211}^{V,\textrm{LO}}]&=&-\frac{\alpha_{e}}{\pi M_Z^2}\frac{Q}{M_N}\Big[\frac{1}{4}\log\frac{\nu+Q^2}{\nu-Q^2}+\frac{2M_NQ+\delta}{8M_N^2}\log\frac{4M_N^4}{\nu^2-Q^4}\Big],\nonumber\\
{\color{black}\textrm{Re}[C_{\gamma Z,212}^{V,\textrm{LO}}]}&=&\color{black}-\frac{\alpha_{e}}{\pi
M_Z^2}\frac{Q}{M_N}\Big[\frac{1}{2}\pi^2+\frac{7}{8}\log\frac{\nu+Q^2}{\nu-Q^2}+\frac{1}{2}R_{\textrm{IR}}\Big],\nonumber\\
\textrm{Re}[C_{\gamma Z,221}^{V,\textrm{LO}}]&=& \frac{\alpha_{e}}{\pi M_Z^2 }\frac{Q}{M_N}\Big[\frac{1}{8}\log\frac{\nu+Q^2}{\nu-Q^2}-\frac{2M_NQ+\delta}{16M_N^2}\log\frac{4M_N^4}{\nu^2-Q^4}
\Big], \nonumber\\
\textrm{Re}[C_{\gamma Z,222}^{V,\textrm{LO}}]&=& \frac{\alpha_e}{\pi M_Z^4}\frac{Q(2M_NQ+\delta)}{144M_N}\Big[7-12\log\frac{M_Z}{M_N}\Big], \nonumber\\
\textrm{Re}[C_{\gamma Z,311}^{V,\textrm{LO}}]&=& \frac{\alpha_{e}}{\pi M_Z^2}\frac{M_N}{Q}\Big[\frac{5}{4}+3\log\frac{M_Z}{M_N}\Big], \nonumber\\
\textrm{Re}[C_{\gamma Z,312}^{V,\textrm{LO}}]&=&\frac{\alpha_{e}}{\pi M_Z^2 }\frac{M_N}{Q}\Big[\frac{1}{2}+2\log\frac{M_Z}{M_N}\Big],\nonumber\\
\textrm{Re}[C_{\gamma Z,321}^{V,\textrm{LO}}]&=&\textrm{Re}[C_{\gamma Z,312}^{V,\textrm{LO}}],\nonumber\\
\textrm{Re}[C_{\gamma Z,322}^{V,\textrm{LO}}]&=&-\frac{\alpha_{e}}{\pi M_Z^2 }\frac{M_N}{Q}\Big[\frac{1}{4}+\frac{9M_Z^2}{16M_N^2}-\log\frac{M_Z}{M_N}+\frac{3M_Z^2}{8M_N^2}R_{\textrm{UV}}\Big],
\label{eq-low-energy-expressions-Re-CV}
\end{eqnarray}
and
\begin{eqnarray}
\textrm{Re}[C_{\gamma Z,113}^{A,\textrm{LO}}]&=&-\frac{\alpha_{e}}{\pi M_Z^2}\frac{2M_NQ+\delta}{Q^2}\Big[\frac{9}{4}+3\log\frac{M_Z}{M_N}+\frac{Q^2}{2M_NQ+\delta}R_{\textrm{IR}}\Big], \nonumber\\
\textrm{Re}[C_{\gamma Z,123}^{A,\textrm{LO}}]&=& -\frac{\alpha_{e}}{\pi M_Z^2}\frac{2M_NQ+\delta}{Q^2}\Big[\frac{9}{4}+3\log\frac{M_Z}{M_N}\Big], \nonumber\\
\textrm{Re}[C_{\gamma Z,213}^{A,\textrm{LO}}]&=&\frac{\alpha_{e}}{\pi M_Z^2}\frac{M_N}{Q}\Big[\frac{9}{2}+6\log\frac{M_Z}{M_N}\Big],\nonumber\\
\textrm{Re}[C_{\gamma Z,223}^{A,\textrm{LO}}]&=& \textrm{Re}[C_{\gamma Z,213}^{A,\textrm{LO}}],\nonumber\\
\textrm{Re}[C_{\gamma Z,313}^{A,\textrm{LO}}]&=&\frac{\alpha_{e}}{\pi M_Z^2 }\frac{M_N}{Q}\Big[\frac{9}{4}+3\log\frac{M_Z}{M_N}\Big],\nonumber\\
\textrm{Re}[C_{\gamma Z,323}^{A,\textrm{LO}}]&=& \textrm{Re}[C_{\gamma Z,313}^{A,\textrm{LO}}],
\label{eq-low-energy-expressions-Re-CA}
\end{eqnarray}
where $\alpha_{e}$ is the fine structure constant and
\begin{eqnarray}
\textrm{R}_{\textrm{IR}}&=&\log\frac{\nu+Q^2}{\nu-Q^2}\Big(\log\frac{4M_N^2\bar{\mu}_{\textrm{IR}}^2}{\nu^2-Q^4}
+\frac{1}{\widetilde{\epsilon}_{\textrm{IR}}}\Big), \nonumber\\
\textrm{R}_{\textrm{UV}}&=&\log\frac{\bar{\mu}_{\textrm{UV}}^2}{M_Z^2}+\frac{1}{\widetilde{\epsilon}_{\textrm{UV}}},\nonumber\\
\frac{1}{\widetilde{\epsilon}_{\textrm{IR},\textrm{UV}}}&=&\frac{1}{\epsilon_{\textrm{IR},\textrm{UV}}}-\gamma_E+\ln4\pi.
\end{eqnarray}
We would like to mention that the nonzero NLO contributions are also present in $\textrm{Re}[C_{\gamma Z,211}^{V,\textrm{LO}}]$ and $\textrm{Re}[C_{\gamma Z,221}^{V,\textrm{LO}}]$ due to the small factor $\log\frac{\nu+Q^2}{\nu-Q^2}$ in the LO contributions. At first glance, the form of $\textrm{Re}[C_{\gamma Z,212}^{V,\textrm{LO}}]$ appears significantly different from the other terms. However, in our practical calculations, we have verified this form using independent numerical results obtained from the LoopTools package, and we have found that the two results are consistent.

The nonzero LO contributions $\textrm{Im}[C_{\gamma Z,ijk}^{V,\textrm{LO}}]$ and $\textrm{Im}[C_{\gamma Z,ijk}^{A,\textrm{LO}}]$ are expressed as
\begin{eqnarray}
\textrm{Im}[C_{\gamma Z,111}^{V,\textrm{LO}}]&=& -\frac{\alpha_{e}}{M_Z^2}\Big[\frac{(4M_NQ+\delta)\delta}{4 M_N^2Q^2}-I_{\textrm{IR}}
\Big], \nonumber\\
\textrm{Im}[C_{\gamma Z,112}^{V,\textrm{LO}}]&=&-\frac{\alpha_{e}}{M_Z^2 }\frac{(2M_NQ+\delta)^2}{8 M_N^2Q^2},  \nonumber\\
\textrm{Im}[C_{\gamma Z,121}^{V,\textrm{LO}}]&=&\textrm{Im}[C_{\gamma Z,112}^{V,\textrm{LO}}], \nonumber\\
\textrm{Im}[C_{\gamma Z,122}^{V,\textrm{LO}}]&=&-\frac{\alpha_{e}}{M_Z^2 }\frac{(M_NQ+\delta)(2M_NQ+\delta)(3M_NQ+\delta)}{16 M_N^4Q^2}, \nonumber\\
\textrm{Im}[C_{\gamma Z,211}^{V,\textrm{LO}}]&=&\frac{\alpha_{e}}{ M_Z^2}\frac{Q}{4M_N},\nonumber\\
\textrm{Im}[C_{\gamma Z,212}^{V,\textrm{LO}}]&=&\frac{\alpha_{e}}{M_Z^2}\frac{Q}{8M_N}
\Big[7+4I_{\textrm{IR}}\Big], \nonumber\\
\textrm{Im}[C_{\gamma Z,221}^{V,\textrm{LO}}]&=&-\frac{\alpha_{e}}{M_Z^2 }\frac{Q}{8 M_N}, \nonumber\\
\textrm{Im}[C_{\gamma Z,222}^{V,\textrm{LO}}]&=&-\frac{\alpha_{e}}{M_Z^4 }\frac{Q(2M_NQ+\delta)^3}{192 M_N^5}, \nonumber\\
\textrm{Im}[C_{\gamma Z,311}^{V,\textrm{LO}}]&=&\frac{\alpha_{e}}{ M_Z^2}\frac{2M_NQ+\delta}{4M_NQ},\nonumber\\
\textrm{Im}[C_{\gamma Z,312}^{V,\textrm{LO}}]&=&\frac{\alpha_{e}}{M_Z^2 }\frac{2M_NQ+\delta}{8M_NQ},\nonumber\\
\textrm{Im}[C_{\gamma Z,321}^{V,\textrm{LO}}]&=&\textrm{Im}[C_{\gamma Z,312}^{V,\textrm{LO}}],\nonumber\\
\textrm{Im}[C_{\gamma Z,322}^{V,\textrm{LO}}]&=&\frac{\alpha_{e}}{M_Z^2 }\frac{(M_NQ+\delta)(3M_NQ+\delta)}{16 M_N^3Q},
\label{eq-low-energy-expressions-Im-CV}
\end{eqnarray}
and
\begin{eqnarray}
\textrm{Im}[C_{\gamma Z,113}^{A,\textrm{LO}}]&=&\frac{\alpha_{e}}{ M_Z^2}\Big[-\frac{2M_N^2Q^2+4M_NQ\delta+\delta^2}{2M_N^2Q^2}
+I_{\textrm{IR}} \Big], \nonumber\\
\textrm{Im}[C_{\gamma Z,123}^{A,\textrm{LO}}]&=&-\frac{\alpha_{e}}{M_Z^2}\frac{(2M_NQ+\delta)^2}{2M_N^2Q^2}, \nonumber\\
\textrm{Im}[C_{\gamma Z,213}^{A,\textrm{LO}}]&=&\frac{\alpha_{e}}{M_Z^2}\frac{2M_NQ+\delta}{M_NQ},\nonumber\\
\textrm{Im}[C_{\gamma Z,223}^{A,\textrm{LO}}]&=&\textrm{Im}[C_{\gamma Z,213}^{A,\textrm{LO}}],\nonumber\\
\textrm{Im}[C_{\gamma Z,313}^{A,\textrm{LO}}]&=&\frac{\alpha_{e}}{M_Z^2}\frac{2M_NQ+\delta}{2M_NQ},\nonumber\\
\textrm{Im}[C_{\gamma Z,323}^{A,\textrm{LO}}]&=&\textrm{Im}[C_{\gamma Z,313}^{A,\textrm{LO}}]
\label{eq-low-energy-expressions-Im-CA}
\end{eqnarray}
with
\begin{eqnarray}
I_{\textrm{IR}}&\equiv&2\log\frac{2M_N\bar{\mu}_{\textrm{IR}}}{\nu+Q^2}+\frac{1}{\widetilde{\epsilon}_{\textrm{IR}}}.
\end{eqnarray}

\subsection{Expressions for ${\cal M}_{\gamma Z,jk}^{\pm\pm\pm\pm,X,\textrm{LO}}$}
\setcounter{equation}{0}
\renewcommand\theequation{B.\arabic{equation}}

The nonzero LO contributions ${\cal M}_{\gamma Z,jk}^{++++,V,\textrm{LO}}$ and ${\cal M}_{\gamma Z,j3}^{++++,A,\textrm{LO}}$ are expressed as
\begin{eqnarray}
\textrm{Re}[{\cal M}_{\gamma Z,11}^{++++,V,\textrm{LO}}]&=&\frac{\alpha_e h}{4\pi}
\Big[(8M_N^2Q^2+z^2)(5+12\log\frac{M_Z}{M_N})+4z^2(\pi^2+\log\frac{\nu+Q^2}{\nu-Q^2}+R_{\textrm{IR}})\Big],\nonumber\\
\textrm{Re}[{\cal M}_{\gamma Z,12}^{++++,V,\textrm{LO}}]&=&\frac{\alpha_e h}{2\pi}
\Big[(8M_N^2Q^2+z^2)(1+4\log\frac{M_Z}{M_N})+4(2M_NQ+\delta)Q^2R_{\textrm{IR}}\Big],\nonumber\\
\textrm{Re}[{\cal M}_{\gamma Z,21}^{++++,V,\textrm{LO}}]&=&\frac{\alpha_e h}{2\pi}
(8M_N^2Q^2+z^2)(1+4\log\frac{M_Z}{M_N}),\nonumber\\
\textrm{Re}[{\cal M}_{\gamma Z,22}^{++++,V,\textrm{LO}}]&=&-\frac{\alpha_e h}{16\pi }
(8M_N^2Q^2+z^2)\Big[4+\frac{9M_Z^2}{M_N^2}-16\log\frac{M_Z}{M_N}
+\frac{6M_Z^2}{M_N^2}R_{\textrm{UV}}\Big], \nonumber\\
\textrm{Re}[{\cal M}_{\gamma Z,13}^{++++,A,\textrm{LO}}]&=&\frac{\alpha_eh}{\pi}
\Big[6M_N^2(2M_NQ+\delta)(3+4\log\frac{M_Z}{M_N})+(8M_N^2Q^2+z^2)R_{\textrm{IR}}\Big],\nonumber\\
\textrm{Re}[{\cal M}_{\gamma Z,23}^{++++,A,\textrm{LO}}]&=&\frac{6\alpha_e h}{\pi}M_N^2(2M_NQ+\delta)(3+4\log\frac{M_Z}{M_N}),
\end{eqnarray}
and
\begin{eqnarray}
\textrm{Im}[{\cal M}_{\gamma Z,11}^{++++,V,\textrm{LO}}]&=&-\alpha_e hz^2\Big[1+I_{\textrm{IR}}\Big],\nonumber\\
\textrm{Im}[{\cal M}_{\gamma Z,12}^{++++,V,\textrm{LO}}]&=&-\alpha_e h\frac{2M_NQ+\delta}{8M_N^2}
\Big[(20M_N^2Q^2-z^2)+16M_N^2Q^2I_{\textrm{IR}}\Big],\nonumber\\
\textrm{Im}[{\cal M}_{\gamma Z,21}^{++++,V,\textrm{LO}}]&=&\alpha_e h \frac{2M_NQ+\delta}{8M_N^2}
(12M_N^2Q^2+z^2),\nonumber\\
\textrm{Im}[{\cal M}_{\gamma Z,22}^{++++,V,\textrm{LO}}]&=&\alpha_e h \frac{(2M_NQ+\delta)^2}{32M_N^4}
(12M_N^2Q^2+5z^2),\nonumber\\
\textrm{Im}[{\cal M}_{\gamma Z,13}^{++++,A,\textrm{LO}}]&=&\alpha_e h
\Big[(8M_N^2Q^2+3z^2)-(8M_N^2Q^2+z^2)I_{\textrm{IR}}\Big],\nonumber\\
\textrm{Im}[{\cal M}_{\gamma Z,23}^{++++,A,\textrm{LO}}]&=&4\alpha_e h (2M_NQ+\delta)^2.
\end{eqnarray}

The nonzero LO contributions ${\cal M}_{\gamma Z,jk}^{+++-,V,\textrm{LO}}$ and ${\cal M}_{\gamma Z,j3}^{+++-,A,\textrm{LO}}$ are expressed as follows:
\begin{eqnarray}
\textrm{Re}[{\cal M}_{\gamma Z,11}^{+++-,V,\textrm{LO}}]&=&-\frac{\alpha_e h z}{2\pi}M_NQ
\Big[4\pi^2-5+4\log\frac{M_N^3(\nu+Q^2)}{M_Z^3(\nu-Q^2)}+4R_{\textrm{IR}}\Big]\nonumber\\
\textrm{Re}[{\cal M}_{\gamma Z,12}^{+++-,V,\textrm{LO}}]&=&\frac{\alpha_e h z }{\pi}M_NQ
\Big[1+4\log\frac{M_Z}{M_N}+\frac{2M_NQ+\delta}{2M_N}R_{\textrm{IR}} \Big],\nonumber\\
\textrm{Re}[{\cal M}_{\gamma Z,21}^{+++-,V,\textrm{LO}}]&=&\frac{\alpha_e h z}{\pi }M_NQ
\Big[1+4\log\frac{M_Z}{M_N} \Big],\nonumber\\
\textrm{Re}[{\cal M}_{\gamma Z,22}^{+++-,V,\textrm{LO}}]&=&-\frac{\alpha_e h z}{8\pi}M_NQ
\Big[4+\frac{9M_Z^2}{M_N^2}-16\log\frac{M_Z}{M_N}+\frac{6M_Z^2}{M_N^2}R_{\textrm{UV}}\Big],\nonumber\\
\textrm{Re}[{\cal M}_{\gamma Z,13}^{+++-,A,\textrm{LO}}]&=&-\frac{3\alpha_e h z}{\pi }\frac{M_N(2M_NQ+\delta)}{Q}
\Big[3+4\log\frac{M_Z}{M_N}+\frac{2}{3}\frac{Q^2}{2M_NQ+\delta}R_{\textrm{IR}}\Big],\nonumber\\
\textrm{Re}[{\cal M}_{\gamma Z,23}^{+++-,A,\textrm{LO}}]&=&-\frac{3\alpha_e h z}{\pi}\frac{M_N(2M_NQ+\delta)}{Q}
\Big[3+4\log\frac{M_Z}{M_N}\Big],
\end{eqnarray}
and
\begin{eqnarray}
\textrm{Im}[{\cal M}_{\gamma Z,11}^{+++-,V,\textrm{LO}}]&=&2\alpha_e h z M_NQ\Big[1+I_{\textrm{IR}}],\nonumber\\
\textrm{Im}[{\cal M}_{\gamma Z,12}^{+++-,V,\textrm{LO}}]&=&-\alpha_e h z\frac{(2M_NQ+\delta)Q}{8M_N}
\Big[5+4I_{\textrm{IR}}\Big],\nonumber\\
\textrm{Im}[{\cal M}_{\gamma Z,21}^{+++-,V,\textrm{LO}}]&=&3\alpha_e h z \frac{(2M_NQ+\delta)Q}{8M_N},\nonumber\\
\textrm{Im}[{\cal M}_{\gamma Z,22}^{+++-,V,\textrm{LO}}]&=&-\alpha_e h z \frac{ (2M_NQ+\delta)^2Q}{16M_N^3},\nonumber \\
\textrm{Im}[{\cal M}_{\gamma Z,13}^{+++-,A,\textrm{LO}}]&=&-2\alpha_e h z
\frac{(3M_NQ+\delta)(M_NQ+\delta)}{M_NQ}\Big[1-\frac{M_N^2Q^2}{(3M_NQ+\delta)(M_NQ+\delta)}I_{\textrm{IR}}\Big],\nonumber\\
\textrm{Im}[{\cal M}_{\gamma Z,23}^{+++-,A,\textrm{LO}}]&=&-2\alpha_eh z \frac{(2M_NQ+\delta)^2}{M_NQ}.
\end{eqnarray}

The nonzero LO contributions ${\cal M}_{\gamma Z,jk}^{++--,V,\textrm{LO}}$ and ${\cal M}_{\gamma Z,j3}^{++--,A,\textrm{LO}}$ are expressed as
\begin{eqnarray}
\textrm{Re}[{\cal M}_{\gamma Z,11}^{++--,V,\textrm{LO}}]&=&\frac{\alpha_e}{4\pi} hz^2
\Big[4\pi^2-5+4\log\frac{M_N^3(\nu+Q^2)}{M_Z^3(\nu-Q^2)}+4R_{\textrm{IR}}\Big],\nonumber\\
\textrm{Re}[{\cal M}_{\gamma Z,12}^{++--,V,\textrm{LO}}]&=&-\frac{\alpha_e }{2\pi}hz^2
\Big[1+4\log\frac{M_Z}{M_N} \Big],\nonumber\\
\textrm{Re}[{\cal M}_{\gamma Z,21}^{++--,V,\textrm{LO}}]&=&\textrm{Re}[{\cal M}_{\gamma Z,12}^{++--,V,\textrm{LO}}],\nonumber\\
\textrm{Re}[{\cal M}_{\gamma Z,22}^{++--,V,\textrm{LO}}]&=&\frac{\alpha_e }{16\pi} hz^2
\Big[4+\frac{9M_Z^2}{M_N^2}-16\log\frac{M_Z}{M_N}+\frac{6M_Z^2}{M_N^2}R_{\textrm{UV}}\Big],\nonumber\\
\textrm{Re}[{\cal M}_{\gamma Z,13}^{++--,A,\textrm{LO}}]&=&-\frac{\alpha_e}{4\pi} hz^2
\Big[4\pi^2-9+4\log\frac{M_N^3(\nu+Q^2)}{M_Z^3(\nu-Q^2)}+4R_{\textrm{IR}}\Big],\nonumber\\
\textrm{Re}[{\cal M}_{\gamma Z,23}^{++--,A,\textrm{LO}}]&=&\frac{3\alpha_e}{4\pi} hz^2
\Big[3+4\log\frac{M_Z}{M_N}\Big],
\end{eqnarray}
and
\begin{eqnarray}
\textrm{Im}[{\cal M}_{\gamma Z,11}^{++--,V,\textrm{LO}}]&=&-\alpha_eh z^2\Big[1+I_{\textrm{IR}}\Big],\nonumber\\
\textrm{Im}[{\cal M}_{\gamma Z,12}^{++--,V,\textrm{LO}}]&=&-\alpha_e h z^2 \frac{(2M_NQ+\delta)}{8M_N^2},\nonumber\\
\textrm{Im}[{\cal M}_{\gamma Z,21}^{++--,V,\textrm{LO}}]&=&\textrm{Im}[{\cal M}_{\gamma Z,12}^{++--,V,\textrm{LO}}],\nonumber\\
\textrm{Im}[{\cal M}_{\gamma Z,22}^{++--,V,\textrm{LO}}]&=&\alpha_e h z^2 \frac{(2M_NQ+\delta)^2}{32M_N^4},\nonumber\\
\textrm{Im}[{\cal M}_{\gamma Z,13}^{++--,A,\textrm{LO}}]&=&-\textrm{Im}[{\cal M}_{\gamma Z,11}^{++--,V,\textrm{LO}}],\nonumber\\
\textrm{Im}[{\cal M}_{\gamma Z,23}^{++--,A,\textrm{LO}}]&=&-\textrm{Im}[{\cal M}_{\gamma Z,12}^{++--,V,\textrm{LO}}].
\end{eqnarray}

\subsection{Expressions for ${\cal A}_{\gamma Z,ijk}^{X,\textrm{LO}}$}

\setcounter{equation}{0}
\renewcommand\theequation{C.\arabic{equation}}

The nonzero LO contributions $\textrm{Re}[{\cal A}_{\gamma Z,ijk}^{X,\textrm{LO}}]$ are expressed as follows:
\begin{eqnarray}
\textrm{Re}[{\cal A}_{\gamma Z,111}^{V,\textrm{LO}}]&=&-\frac{8\alpha_{e}}{\pi M_Z^2 }M_N^2 Q^2z^2\Big[\pi^2+\log\frac{\nu+Q^2}{\nu-Q^2}+R_{\textrm{IR}}\Big], \nonumber\\
\textrm{Re}[{\cal A}_{\gamma Z,112}^{V,\textrm{LO}}]&=&-\frac{2\alpha_{e}}{\pi M_Z^2}Q^2
\Big[(2M_NQ+\delta)^3+16M_N^2Q^2(2M_NQ+\delta)\log\frac{M_Z}{M_N}+
8M_N^2Q^4R_{\textrm{IR}}\Big],\nonumber\\
\textrm{Re}[{\cal A}_{\gamma Z,121}^{V,\textrm{LO}}]&=&-\frac{2\alpha_{e}}{\pi M_Z^2}Q^2(2M_NQ+\delta)
\Big[(2M_NQ+\delta)^2+16M_N^2Q^2\log\frac{M_Z}{M_N}\Big],\nonumber\\
\textrm{Re}[{\cal A}_{\gamma Z,122}^{V,\textrm{LO}}]&=&\frac{\alpha_{e}}{2\pi M_Z^2}Q^2(2M_NQ+\delta)
\Big[2(4M_N^2+9M_Z^2)Q^2-7z^2\nonumber\\
&&~~-4(8M_N^2Q^2+z^2)\log\frac{M_Z}{M_N}+12M_Z^2Q^2R_{\textrm{UV}}\Big],\nonumber\\
\textrm{Re}[{\cal A}_{\gamma Z,211}^{V,\textrm{LO}}]&=&-\frac{4\alpha_{e}}{\pi M_Z^2}M_N^2Q^4(2M_NQ+\delta)\Big[5+12\log\frac{M_Z}{M_N}+\frac{4Q^2}{2M_NQ+\delta}R_{\textrm{IR}}\Big],\nonumber\\
\textrm{Re}[{\cal A}_{\gamma Z,212}^{V,\textrm{LO}}]&=&-\frac{2\alpha_{e}}{\pi M_Z^2}Q^4
\Big[4M_N^2(2M_NQ+\delta)(1+4\log\frac{M_Z}{M_N})
+(8M_N^2Q^2+z^2)R_{\textrm{IR}}\Big],\nonumber\\
\textrm{Re}[{\cal A}_{\gamma Z,221}^{V,\textrm{LO}}]&=&-\frac{8\alpha_{e}}{\pi M_Z^2}M_N^2Q^4(2M_NQ+\delta)(1+4\log\frac{M_Z}{M_N}), \nonumber\\
\textrm{Re}[{\cal A}_{\gamma Z,222}^{V,\textrm{LO}}]&=& \frac{\alpha_{e}}{\pi M_Z^2}Q^4(2M_NQ+\delta)\Big[4M_N^2+9M_Z^2-16M_N^2\log\frac{M_Z}{M_N}+6M_Z^2R_{\textrm{UV}}\Big],\nonumber\\
\textrm{Re}[{\cal A}_{\gamma Z,113}^{A,\textrm{LO}}]&=&-\frac{2\alpha_{e}}{\pi M_Z^2 }M_N^2 Q^2\Big[3z^2(3+4\log\frac{M_Z}{M_N})
+8Q^2(2M_NQ+\delta)R_{\textrm{IR}}\Big], \nonumber\\
\textrm{Re}[{\cal A}_{\gamma Z,123}^{A,\textrm{LO}}]&=&-\frac{6\alpha_{e}}{\pi M_Z^2 }M_N^2 Q^2z^2(3+4\log\frac{M_Z}{M_N}),\nonumber\\
\textrm{Re}[{\cal A}_{\gamma Z,213}^{A,\textrm{LO}}]&=&-\frac{16\alpha_{e}}{\pi M_Z^2}M_N^2 Q^4(2M_NQ+\delta)\Big[\pi^2+\log\frac{\nu+Q^2}{\nu-Q^2} +R_{\textrm{IR}}\Big],\nonumber\\
\textrm{Re}[{\cal A}_{\gamma Z,223}^{A,\textrm{LO}}]&=&-\frac{\alpha_{e}}{2\pi M_Z^2}Q^4\Big[
8M_N^2Q^2(3+10\log2)+2z^2(-5+11\log2)\nonumber\\
&&~~~~~~~~+8(4M_N^2Q^2-z^2)\log\frac{M_Z}{M_N}+(40M_N^2Q^2+11z^2)\log\frac{M_N^4}{\nu^2-Q^4}\Big].
\label{eq-expression-of-Re-Hijk}
\end{eqnarray}

\end{document}